\RequirePackage{ifpdf}
\documentclass[letterpaper]{article} 
\usepackage[usenames,dvipsnames]{pstricks}
\usepackage{./auxi/jheppub}
\usepackage[english]{babel}
\usepackage[babel]{csquotes}
\usepackage{amsmath}
\usepackage{amsfonts}
\usepackage{amssymb}
\usepackage{mathtools,colonequals}
\usepackage{mathrsfs}
\usepackage{bm}
\usepackage{bbold}
\usepackage{braket}
\usepackage{graphicx}
\usepackage{tikz}
\usetikzlibrary{decorations.pathmorphing}
\usetikzlibrary{decorations.markings}
\usepgflibrary{shapes.geometric}
\usepackage{booktabs}
\usepackage{array}
\usepackage{color}
\usepackage{subcaption}
\usepackage[percent]{overpic}
\usepackage{multirow}

\newcommand{\al}{\alpha}

\newcommand{\de}{\delta}

\newcommand{\ep}{\epsilon}

\renewcommand{\th}{\theta}

\newcommand{\m}{\mu}
\newcommand{\n}{\nu}
\renewcommand{\r}{\rho}
\newcommand{\si}{\sigma}

\renewcommand{\O}{\Omega}

\newcommand{\pa}{\partial}

\newcommand{\beq}{\begin{equation}}
\newcommand{\eeq}{\end{equation}}

\newcommand{\bz}{\bar{z}}
\newcommand{\bh}{\bar{h}}
\newcommand{\bp}{\bar{p}}
\newcommand{\Eop}{\mathcal{E}}
\newcommand{\bEop}{\mathcal{\bar{E}}}
\newcommand{\bT}{\overline{T}}
\newcommand{\quellaT}{\mathcal{T}_{\text{\cite{Quella:2006de}}}}
\newcommand{\quellaR}{\mathcal{R}_{\text{\cite{Quella:2006de}}}}
\newcommand{\ourT}{\mathcal{T}}
\newcommand{\ourR}{\mathcal{R}}
\newcommand{\braketi}[1]{\braket{#1}_\textup{I}}
\newcommand{\keti}[1]{\ket{#1}_\textup{I}}

\newcommand{\ii}{\mathrm{i}}

\def\Im{{\mathcal{I}}}

\def\w{\omega}
\def\wb{\bar{\omega}}

\def\w{\omega}
\def\wb{\bar{\omega}}
\def\r{\rho}
\def\rb{\bar{\rho}}

\newcommand{\rhob}{\bar{\rho}}

\usepackage{xargs}
\newcommandx\Fbraket[1][usedefault, addprefix=\global, 1=]{\braket{#1}}%
\newcommandx\Fket[1][usedefault, addprefix=\global, 1=]{\ket{#1}}%
\newcommandx\FNordered[1][usedefault, addprefix=\global, 1=]{\,:\!{#1}\!:\,}%

\author{Marco Meineri,}
\author{Joao Penedones,}
\author{Antonin Rousset}

\affiliation{Fields and Strings Laboratory, Institute of Physics, \'{E}cole Polytechnique F\'{e}d\'{e}rale de Lausanne (EPFL)\\
Rte de la Sorge, BSP 728, CH-1015 Lausanne, Switzerland}

\emailAdd{marco.meineri@gmail.com, jpenedones@gmail.com, antonin@rousset.com}

\bigskip
\abstract{
We set up a scattering experiment of matter against an impurity which separates two generic one-dimensional critical quantum systems. We compute the flux of reflected and transmitted energy, thus defining a precise measure of the transparency of the interface between the related two-dimensional conformal field theories. If the largest symmetry algebra is Virasoro, we find that the reflection and transmission coefficients are independent of the details of the initial state, and are fixed in terms of the central charges and of the two-point function of the displacement operator. The situation is more elaborate when extended symmetries are present. Positivity of the total energy flux at infinity imposes bounds on the coefficient of the two-point function of the displacement operator, which controls the free-energy cost of a small deformation of the interface. Finally, we study out-of-equilibrium steady states of a critical system connecting two reservoirs at different temperatures. In the absence of extended symmetries, our result implies that the energy flux across an impurity is proportional to the difference of the squared temperatures and controlled by the reflection coefficient. \phantom{\cite{Quella:2006de}}}

\title{Colliders and conformal interfaces}


\begin{document}

\maketitle
\newpage
\section{Introduction}

Although the Euclidean and the Lorentzian formulations of a quantum field theory (QFT) are in principle equivalent \cite{osterwalder1973}, very different properties of the theory are accessible in one or the other setup. This is true in particular for a conformal field theory (CFT), as it has become more and more apparent in recent years. The $a$-theorem \cite{Komargodski:2011vj}, the conformal collider bounds \cite{Hofman:2008ar,Hofman:2016awc} the averaged null energy condition (ANEC) \cite{Faulkner:2016mzt,Hartman:2016lgu}, the analytic structure constraining the spectrum of every CFT \cite{Komargodski:2012ek,Fitzpatrick:2012yx,Alday:2016njk,Caron-Huot:2017vep,Kravchuk:2018htv}, all these results have been obtained by exploiting the way in which unitarity and causality are encoded in the Lorentzian regime of the theory. In this work, we study the real time dynamics of a CFT in the presence of an interface. This has two complementary purposes. On the one hand, real time evolution allows to probe the conformal interface via a scattering process, whose associated observables provide qualitative information about the interface, and constraints on the attached CFT data. On the other hand, the interface itself, like any other operator, is essentially a probe of the CFT: by scattering conformal matter against an interface, we also gather information on the conformal matter itself. For instance, one can excite the vacuum with different local operators, and study the dependence of the scattering observables on the initial state. 
In fact, interfaces are special probes, since in general they glue together two different CFTs. One might hope that, measuring the transparency of the set of conformal interfaces, it is possible to learn general facts about the space of CFTs \cite{Douglas:2010ic}.

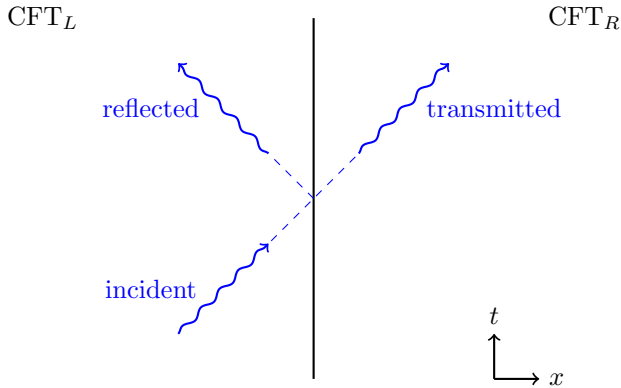
\begin{figure}[b]
\centering
\begin{tikzpicture}[scale=1.2]
\draw [thick]  (0,-2) -- (0,2);
\draw [thick,->] (2,-2) -- (2.5,-2) node[anchor=west] {$x$};
\draw [thick,->] (2,-2) -- (2,-1.5) node[above] {$t$};
\node at (3,2) {CFT$_R$};
\node at (-3,2) {CFT$_L$};
\draw [dashed,blue] (-0.47,-0.47) -- (0.47,0.47);
\draw [dashed,blue] (0,0) -- (-0.47,0.47);
\draw [thick,blue,decorate,decoration={snake,amplitude=1pt},->] (-0.5,0.5) -- (-1.5,1.5);
\node [blue] at (-1.8,1) {reflected};
\draw [thick,blue,decorate,decoration={snake,amplitude=1pt},->] (0.5,0.5) -- (1.5,1.5);
\node [blue] at (2,1) {transmitted};
\draw [thick,blue,decorate,decoration={snake,amplitude=1pt},->] (-1.5,-1.5) -- (-0.5,-0.5);
\node [blue] at (-1.8,-1) {incident};
\end{tikzpicture}
\caption{The interface is the worldline of an impurity placed at $x=0$, which separates a CFT$_L$ from a CFT$_R$.}
\label{fig:interface}
\end{figure}
Our  setup will be two dimensional. The restriction to two dimensions is not conceptually necessary, but practically helpful for explicit calculations. We shall consider a pair of CFTs, denoted CFT$_L$ and CFT$_R$, glued at an interface as depicted in figure \ref{fig:interface}.
We can then prepare an excitation far from the interface and let it propagate until it collides with the interface. We define transmission and reflection coefficients as the fraction of energy that is transmitted or reflected across the interface:
\beq
\mathcal{T} = \frac{{\rm transmitted\  energy} }{\rm{incident\  energy}}\,,\qquad \qquad
\mathcal{R} = \frac{{\rm reflected\  energy} }{\rm{incident\  energy}}\,.
\eeq
Energy conservation implies $\mathcal{T}+\mathcal{R}=1$ and positivity of the total energy transmitted and reflected leads to $\mathcal{T}\ge 0$ and $\mathcal{R}\ge 0$. Clearly, we can also define analogous transmission and reflection coefficients associated to other conserved quantities.
The advantage of  measuring this type of inclusive quantities in CFT was emphasized in \cite{Hofman:2008ar}, where the expectation value of the energy flux operator, or of multiple insertions thereof, was measured in a state created by a local operator. 

A priori we expect $\mathcal{T}$ and $\mathcal{R}$ to depend on what excitations we are shooting at the interface. 
Remarkably, we shall show that in a generic CFT they are completely independent of the details of the incoming excitation. 
More precisely, we will see that they are entirely determined in terms of the central charges $c_L$ and $c_R$ of the left and right CFTs and the two point function of the stress tensor
across the interface,
\beq
\braketi{T_L(z_1)T_R(z_2)}=\frac{c_{LR}/2}{(z_1-z_2)^{4}}~,
\eeq
where $T_L\, (T_R)$ denotes the holomorphic stress tensor of the left (right) CFT.
Our main result, is that for any excitation colliding with the interface the transmission coefficients  are given by
\beq
\ourT_L = \frac{c_{LR}}{c_L}~, \quad \qquad \ourT_R = \frac{c_{LR}}{c_R}~, 
\label{mainresult}
\eeq
where the subscript of $\ourT$ indicates the origin of the incoming excitation.
The combination $\frac{c_{LR}}{c_L+c_R}$ was prophetically termed transmission coefficient in \cite{Quella:2006de}.
Our results show that this combination is indeed a weighted average of the physical energy transmission coefficients $\ourT_L$ and $\ourT_R$.
Furthermore, \eqref{mainresult} leads to the bounds
\beq
0<c_{LR}<\min (c_L,c_R)\,,\qquad
\qquad
0<\ourT_L <\min \left(1 , \frac{c_R}{c_L}\right)\,,\qquad
\qquad
0<\ourT_R <\min \left(1 , \frac{c_L}{c_R}\right)\,.
\label{introbounds}
\eeq 
This shows that it is impossible to fully transmit energy from the CFT with larger central charge to the CFT with smaller central charge. 

The situation changes if the spectrum of one of the CFTs contains more than one spin 2 conserved quasi-primary. This is commonly the case when the theory is endowed with an extended symmetry. In this case, eqs. \eqref{mainresult} and \eqref{introbounds} are only valid, in general, if the excitation is created by the stress tensor. The reflection and transmission coefficients are state dependent, and can be computed from the knowledge of the relevant CFT data.

The paper is organized as follows. Section \ref{sec:propagation} contains preliminary remarks on the propagation of matter in a $2d$ CFT. In section \ref{sec:R&T} we define the reflection and transmission coefficients and derive the main results \eqref{mainresult} and \eqref{introbounds}. In section \ref{sec:examples}, we check and illustrate the results in a few examples. Section \ref{sec:steady} is dedicated to an application of our main result to the physics of non-equilibrium steady states.  Finally, we draw our conclusions and discuss future directions in section \ref{sec:summary}. A few technical details are relegated to appendices.

\section{Propagation of conformal matter}
\label{sec:propagation}

\begin{figure}[]
\centering
\begin{tikzpicture}[scale=1.1]
\draw [thick] (0,-2) -- (2,0);
\draw  [thick] (2,0) -- (0,2);
\draw [thick] (0,2) -- (-2,0);
\draw [thick] (-2,0) -- (0,-2);
\draw [thick,->] (3,-1) -- (3.6,-1.6) node[anchor=west] {$w$};
\draw [thick,->] (3,-1) -- (3.6,-0.4) node[anchor=west] {$\bar{w}$};
\draw [dashed] (-0.5,-1.5) -- (1.5,0.5) node[pos=0.5,below=2] {$\bEop$};
\draw [dashed] (1.5,-0.5) -- (-0.5,1.5) node[pos=0.5,above=2] {$\Eop$};
\end{tikzpicture}
\caption{The Penrose diagram of $2d$ Minkowski space. The coordinates $w,\, \bar{w}$ are related to the $z,\,\bz$ coordinates as $z=\tan w$ and similarly for $\bz$. The dashed lines mark the integration contours of the ANEC operators \eqref{energyOperators}, which in these coordinates read $\Eop=-1/2\pi\int\! dw\, \cos^2\! w\, T(w)$ -- up to the vacuum energy which we subtracted off -- and similarly for $\bEop$. The contours can be freely translated, therefore $\Eop$ measures the total energy -- equal to the $\bar{z}$ component of the momentum -- which reaches infinity on the right, and similarly $\bEop$.}
\label{fig:penrose}
\end{figure}
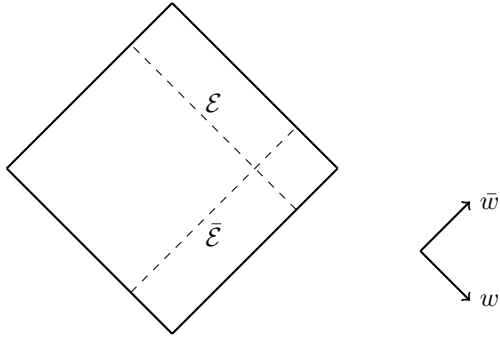

\begin{figure}[h]
\centering
\begin{tikzpicture}[scale=1.2]
\draw  [thick,->] (0,0) -- (3,3) node [anchor=west] {$\bp$};
\draw [thick,->]  (0,0) -- (-3,3) node [anchor=east] {$-p$};
\draw [thick,->]  (0,0) -- (0,3.5) node [anchor=north east] {$p^0$};
\draw [thick,->]  (-3,0) -- (3,0) node [anchor=west] {$p^1$};
\draw [dashed] (-1,1) -- (1,3);
\draw [dashed] (1,3) -- (2,2);
\draw [thick, red, ->] (0,0) --(1,3) node [above]{$p^\mu$};
\draw [thick, red, ->] (2,0) -- (2,2) node [right=8,below=1] {$\frac{\bp}{2}$};
\draw [thick, red, ->] (-1,0) -- (-1,1) node [left=10,below=1] {$-\frac{p}{2}$};
\end{tikzpicture}
\caption{The timelike momentum $p^\mu$ uniquely decomposes in the lightlike momenta of the right and left movers. The time components of the latter coincide with the energy carried to infinity towards the right and towards the left.}
\label{fig:momentumDecomp}
\end{figure}
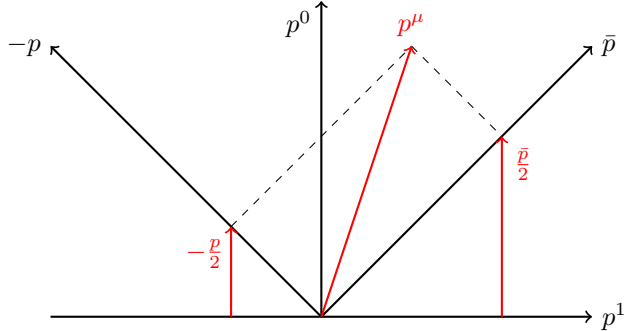

We begin by highlighting a simple consequence of holomorphy on the propagation of conformal matter, which will provide intuition when we construct the initial state for our collider experiment. In a $2d$ CFT, the stress tensor splits into a holomorphic and an antiholomorphic part. In particular, the expectation value of the energy density in any state of the theory splits accordingly:
\beq
\braket{T^{00}(x,t)} = - \frac{1}{2\pi}\left( \braket{T(z)} + \braket{\bT(\bz)} \right)~, \qquad z=x-t~,\ \bz=x+t~.
\label{Tsplit}
\eeq
Our conventions are summarized in appendix \ref{app:conventions}. Eq. \eqref{Tsplit} teaches us that the energy density in any state consists of a (holomorphic) right-moving and a (antiholomorphic) left-moving parts, both propagating rigidly and both at the speed of light. In particular, the left-moving and right-moving wave packets cross each other without exchanging energy. Two interesting observables compute the flux of energy, which equals the flux of spatial momentum, that reaches infinity towards the right or towards the left respectively
\cite{Hofman:2008ar} -- see fig. \ref{fig:penrose}:
\beq
\Eop = -\frac{1}{2\pi} \int_{-\infty}^{+\infty}\! dz\, T(z)~,\qquad 
\bEop =- \frac{1}{2\pi} \int_{-\infty}^{+\infty}\! d\bz\, \overline{T}(\bz)~.
\label{energyOperators}
\eeq
Notice that the operators $\Eop$ and $\bEop$ are ANEC operators. In fact, in $2d$ they simply measure the total momentum $P_\mu$ along the null directions $t\pm x$. Therefore, 
$\Eop=P_z$ and $\bEop=-P_{\bz}$ have non-negative  expectation value in any state.
 We can, for instance, measure the fraction of energy carried away in the two directions in a state created by acting with a local operator in a region of spacetime:
\beq
\ket{O,f}=\int d^2x\, f(x)\, O(x) \ket{0}~,
\label{stateO}
\eeq
where $f(z,\bz)$ is some appropriate wave packet.
The state can be made normalizable by moving the local operator $O$ in imaginary time. The fraction of energy which propagates on average towards the right and the left is computed respectively by the expectation value of $\Eop$ and $\bEop$ in the state \eqref{stateO}, and is therefore completely fixed by conformal invariance and by the choice of wave packet. As an example, let us compute the transmitted and reflected energy in a plane wave state
\beq
\ket{O,p^\mu}=\int d^2x\, e^{\ii p \cdot x}\, O(x) \ket{0}~.
\label{planewaveO}
\eeq
Such a state is actually an eigenstate of the ANEC operators \eqref{energyOperators}:\footnote{This is obtained by projecting eqs. \eqref{epNoInterface} onto a state created by a string of generic local operators. In the resulting correlator, the $\ii \ep$ prescription implies that the contour of integration of the ANEC operators \eqref{energyOperators} separates $O$ from the other insertions. We then close the contour towards the operator $O$, and Fourier transform the result. Notice that only the single pole in the OPE of the stress tensor with the operator $O$ contributes, hence eq. \eqref{epNoInterface} is valid for any linear combination of primaries and descendants.} 
\beq
\Eop\ket{O,p^\mu} = \frac{\bp}{2} \ket{O,p^\mu}~, \qquad
\bEop\ket{O,p^\mu} =- \frac{p}{2} \ket{O,p^\mu}~.
\label{epNoInterface}
\eeq
This is the expected result if the CFT is made of massless carriers of energy and momentum -- see fig. \ref{fig:momentumDecomp}.
It is also instructive to compute the norm of the state \eqref{planewaveO}, which contains the same information as the spectral density of the two-point function of the operator $O$. For simplicity, we choose $O$ to be a scaling operator with dimensions $(h,\bh)$, and denote the corresponding state $\ket{h,\bh,p^\m}$. 
The state is  delta--function normalizable:
\beq
\braket{h,\bh,p^\m|h,\bh,p'^\m} = \frac{(2\pi)^4}{2} \frac{1}{\Gamma(2h)\Gamma(2\bh)} \left(\frac{\bp}{2}\right)^{2h-1}  \left(-\frac{p}{2}\right)^{2\bh-1} \delta^2(p^\m-p'^\m) \theta(p^0-|p^1|)~.
\label{normNoInt}
\eeq
The holomorphic limit must be understood in the sense of distributions:
\beq
\braket{h,0,p^\m|h,0,p'^\m} = \frac{(2\pi)^4}{\Gamma(2h)} (p^0)^{2h-1} \delta^2(p^\m-p'^\m) \delta(p^0-p^1)\theta(p^0)~.
\label{holoNorm}
\eeq
We see that a holomorphic field only creates right moving plane waves.

Let us now consider the situation in the presence of a conformal interface, which we think of as an impurity placed at $x=0$ -- see fig. \ref{fig:interface}. 
Since the holomorphic splitting in eq. \eqref{Tsplit} is a local property, conformal matter still propagates freely until it hits the interface. On the interface, the stress tensor is subject to the following gluing condition:
\beq
T_L(z)+\bT_R(\bz)=\bT_L(\bz) + T_R(z)~, \qquad z+\bz=0~,
\label{Tgluing1}
\eeq
where $T_L\, (T_R)$ is the stress tensor of the CFT on the left (right). We shall come back to this condition in subsection \ref{subsec:dcft}. For now, let us only point out that matter is allowed to be reflected at the interface, but not absorbed.  The perturbations incoming from the left and from the right -- which add up to the left hand side of eq. \eqref{Tgluing1} -- are glued to the outgoing perturbations -- the right hand side of eq. \eqref{Tgluing1}. The amount of energy being reflected and transmitted is not fixed, it depends on the details of the boundary condition and, in principle, on the state. As previewed in the introduction, the amount of transmitted and reflected energy is the observable we would like to focus on in the following. The expectation value of the ANEC operators \eqref{energyOperators} precisely measures these quantities.

As we just saw, matter propagation is only altered by the presence of the interface locally, at $x=0$. On the contrary, the construction of the state is a Euclidean process, strongly influenced by the lack of translational invariance of the vacuum. In fact, the crucial step in our construction will be the definition of an appropriate class of states where reflection and transmission of energy can easily be measured. We do so in the next section, after a brief reminder of some basics of $2d$ defect conformal field theory.

\section{Reflection and transmission coefficients from a collider experiment}
\label{sec:R&T}

In this section, we consider a one dimensional conformal interface which preserves a given (higher spin) symmetry. We define reflection and transmission coefficients, by computing the flux at infinity of the charge associated to this symmetry in a specific class of states. As we review in the next subsection, a conformal interface by definition preserves the maximal possible amount of symmetry generated by the stress tensor, and in particular it conserves energy. Therefore, we focus on the reflection and transmission of energy, whose flux is measured by the operators \eqref{energyOperators}. In subsection \ref{subsec:global}, we consider the case of a global charge.

\subsection{Generalities on defect CFT}
\label{subsec:dcft}

Let us consider an interface between two CFTs in Euclidean signature. The interface runs vertically as in fig. \ref{fig:interface}, but we now replace the coordinate $t$ with the Euclidean counterpart $\tau=\ii t$, and we take $\tau$ to be real. The plane is spanned by the complex coordinate $z=x+\ii \tau$. Imposing that the interface is invariant under $\tau$-translations and scale transformations, we find the following operator equations:\footnote{The integrated operators in eqs. \eqref{int_tran_scale} have vanishing matrix elements between the states created by insertions on the left and on the right of the interface. 
Such insertions generate the full Hilbert spaces on the left and on the right. This is not true if we consider a junction among multiple CFTs rather than an interface. The argument can be easily modified by use of the folding trick. The system can be replaced by a boundary condition for the tensor product of the CFTs. Then, the operator whose integral vanishes in any state is the sum of the appropriate component of the stress tensors of all the CFTs.}
\begin{align}
\int_{-\infty}^\infty\! d\tau\, \big[T_L(z=\ii \tau) - \bT_L(\bz=-\ii \tau)-T_R(z=\ii \tau)+\bT_R(\bz=-\ii \tau)\big]
&=0~, \qquad \\
\int_{-\infty}^\infty\! d\tau\, \tau \big[T_L(z=\ii \tau) - \bT_L(\bz=-\ii \tau)-T_R(z=\ii \tau)+\bT_R(\bz=-\ii \tau)\big]
&=0~.
\label{int_tran_scale}
\end{align}
Notice that the stress tensors are evaluated at the location of the interface: holomorphy and translational invariance ensure that no singularities arise. These equations could be satisfied with a local operator $\th$ on the interface such that
\beq
T_L(z=\ii \tau) - \bT_L(\bz=-\ii \tau)-T_R(z=\ii \tau)+\bT_R(\bz=-\ii \tau)= \partial_\tau^2 \theta(\tau)~.
\eeq
However, $\theta$ has dimension zero, and scale invariance requires its two-point function to be a constant. Then $\braketi{\pa_\tau \theta\, \pa_\tau \theta}=0$, and unitarity of the Hilbert space at constant $\tau$ requires $\partial_{\tau} \theta=0$ \cite{Nakayama:2012ed}.
In other words, there is no defect stress tensor, and we get the gluing condition \cite{Cardy:1984bb}
\beq
T_L(z=\ii \tau) - \bT_L(\bz=-\ii \tau)=T_R(z=\ii \tau)-\bT_R(\bz=-\ii \tau)~,
\quad \tau \in \mathbb{R}~.
\label{gluingT}
\eeq
The gluing condition actually implies that all the infinitesimal conformal transformations which do not displace the interface correspond to conserved charges in radial quantization around a point on the defect. This amounts to one copy of the Virasoro algebra. Actually, the gluing condition \eqref{gluingT} is not sufficient to define a consistent interface. Further conditions follow from the requirement that the partition function on the strip has the correct interpretation of counting defect states, but we will not need the details in this work. 

Let us first discuss two special classes of solutions to eq. \eqref{gluingT}. The first is the set of the topological interfaces, which obey
\beq
T_L(z=\ii \tau)=T_R(z=\ii \tau)~, \qquad \bT_L(\bz=-\ii \tau)=\bT_R(\bz=-\ii \tau)~,\quad \tau \in \mathbb{R}~.
\label{topological}
\eeq
Since the stress tensor is continuous across the interface, two copies of the Virasoro algebra are preserved, and all correlation functions are independent from the position of the interface. The second extreme case is the following:
\beq
T_L(z=\ii \tau)=\bT_L(\bz=-\ii \tau)~, \qquad T_R(z=\ii \tau)=\bT_R(\bz=-\ii \tau)~,\quad \tau \in \mathbb{R}~.
\label{factorizing}
\eeq
Interfaces of this kind are sometimes called factorizing. The two sides are uncorrelated. Indeed, eq. \eqref{factorizing} implies that $\tau$-translations are implemented by independent conserved charges on the left and on the right of the interface, and this is incompatible with a non vanishing connected correlator involving operators both on the left and on the right.

A topological interface can be freely deformed. On the contrary, an infinitesimal deformation of a generic defect is implemented by the insertion of the so called displacement operator, a protected defect operator which is easily seen to be\footnote{The normalization differs by a factor $2\pi$ from the usual choice in higher dimensions \cite{Billo:2016cpy}.}
\beq
\textup{D} =2 ( T_L-T_R)~.
\label{displacement}
\eeq
The coefficient of the two-point function of the displacement, which we call $C_\textup{D}$, measures the free energy cost of an infinitesimal deformation of the interface -- see \emph{e.g.} \cite{Bianchi:2015liz}. Comparing with eq. \eqref{topological}, we indeed see that, in a unitary theory, $C_\textup{D}=0$ implies that the interface is topological. On the other hand, the maximal value for this coefficient is attained by a factorizing interface \cite{Billo:2016cpy}, where $C_\textup{D}=2 (c_L+c_R)$, $c_L$ and $c_R$ being the central charges of the two CFTs. The two-point function of the displacement is simply related to the quantity called reflection coefficient in \cite{Quella:2006de}:
\beq
\quellaR=\frac{C_\textup{D}}{2 (c_L+c_R)}~.
\label{QRWcd}
\eeq
As mentioned in the introduction, we shall provide a precise meaning for $\quellaR$ in terms of reflected energy in an excited state. Hence, eq. \eqref{QRWcd} showcases a nice connection between dynamics and statistical mechanics. A practical consequence of this connection is that the bounds \eqref{introbounds} translate into bounds on $C_\textup{D}$.

There are similar gluing conditions for conserved currents of any spin. However, the gluing of the left and right movers is less constrained. The linear combination in eq. \eqref{gluingT} is fixed by requiring that the preserved symmetries are those which do not deform the defect. Consider instead the spin one case, which is relevant to the discussion of subsection \ref{subsec:global}. Given a set of global symmetry currents $(J^a,\bar{J}^a)$, it makes sense for the interface to preserve the flux of any twisted current $(J,\Omega\bar{J})$, where $\Omega$ is an automorphism of the algebra which leaves the stress tensor invariant. We then obtain the following gluing condition:
\beq
J_L(\ii \tau) + \Omega_L\bar{J}_L(\overline{\ii \tau})=J_R(\ii \tau)+\Omega_R\bar{J}_R(\overline{\ii \tau})~,
\quad \tau \in \mathbb{R}~.
\label{gluingJ}
\eeq

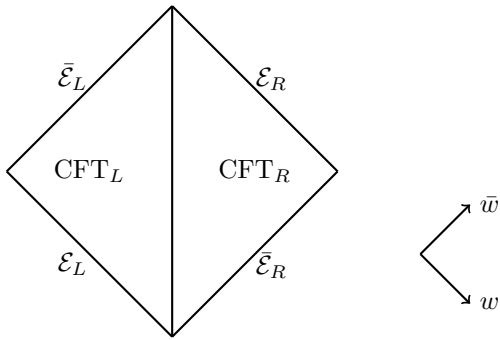
\begin{figure}
\centering
\begin{tikzpicture}[scale=1.1]
\draw [thick] (0,-2) -- (2,0) node[pos=0.6,below=2] {$\bEop_R$};
\draw  [thick] (2,0) -- (0,2) node[pos=0.4,above=2] {$\Eop_R$};
\draw [thick] (0,2) -- (-2,0) node[pos=0.6,above=2] {$\bEop_L$};
\draw [thick] (-2,0) -- (0,-2) node[pos=0.4,below=2] {$\Eop_L$};
\draw [thick] (0,2) -- (0,-2);
\draw [thick,->] (3,-1) -- (3.6,-1.6) node[anchor=west] {$w$};
\draw [thick,->] (3,-1) -- (3.6,-0.4) node[anchor=west] {$\bar{w}$};
\node at (-1,0) {CFT$_L$};
\node at (1,0) {CFT$_R$};
\end{tikzpicture}
\caption{The Penrose diagram associated to the defect CFT. Due to the non trivial gluing conditions, now $\Eop_L$ and $\Eop_R$ do not need to have the same expectation value in a generic state, and similarly for $\bEop_L$ and $\bEop_R$. Energy can be reflected.}
\label{fig:penroseDefect}
\end{figure}

Before ending the subsection, let us point out that the gluing conditions \eqref{gluingT} and \eqref{gluingJ} are easily Wick rotated into eq. \eqref{Tgluing1}. Unless otherwise stated, from now on the stress tensor is always inserted on the Lorentzian slice where the defect is timelike. The Penrose diagram of the defect CFT is then shown in fig. \ref{fig:penroseDefect}, where we also define the energy flux operators $\Eop_L,$ $\Eop_R$ and their antiholomorphic counterparts. They are analogous to the ones in eq. \eqref{energyOperators}, and they are built out of the left and right stress tensors respectively.

\subsection{The state}
\label{subsec:wflim}

\begin{figure}
\centering
\begin{subfigure}{0.31\textwidth}
\begin{overpic}[scale=0.33]{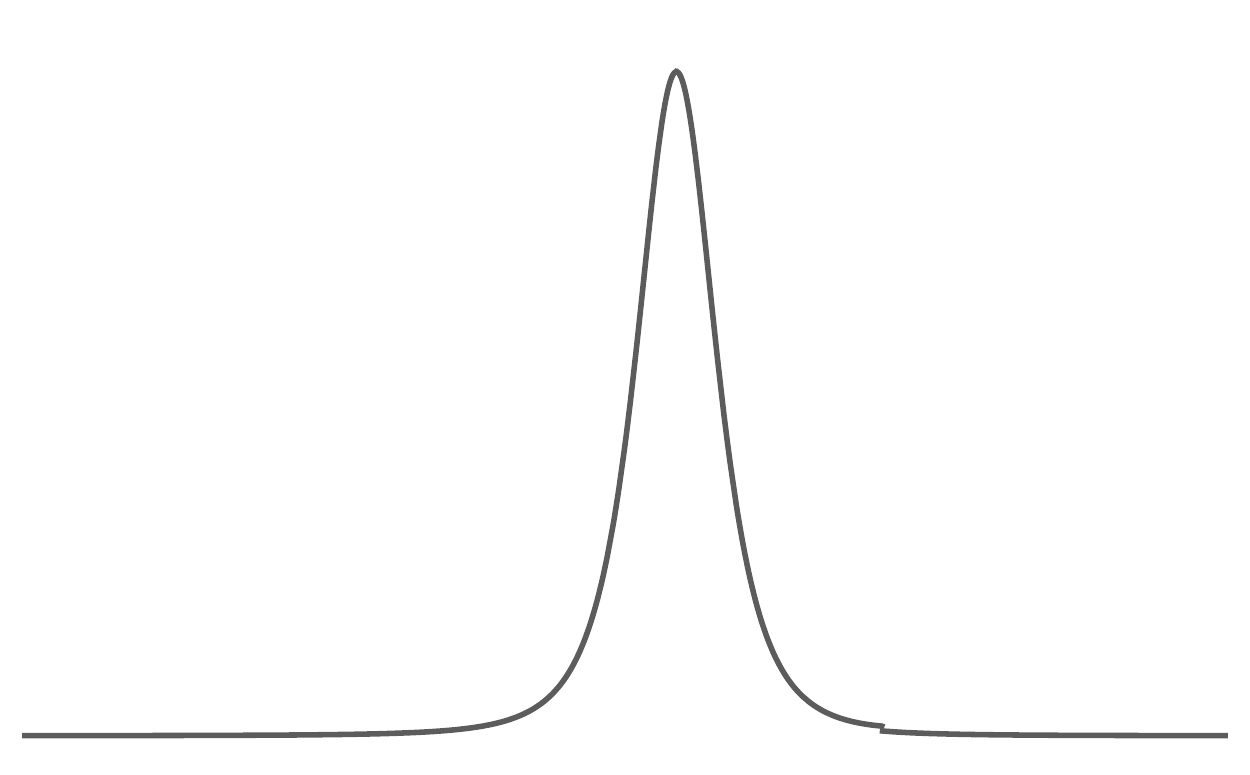}
\put(0,56){$t=0$}
\thicklines
\put(70.8,2.5){\line(0,1){63}}
\put(0,2.5){\line(1,0){103}}
\end{overpic}
\end{subfigure}
\begin{subfigure}{0.31\textwidth}
\begin{overpic}[scale=0.33]{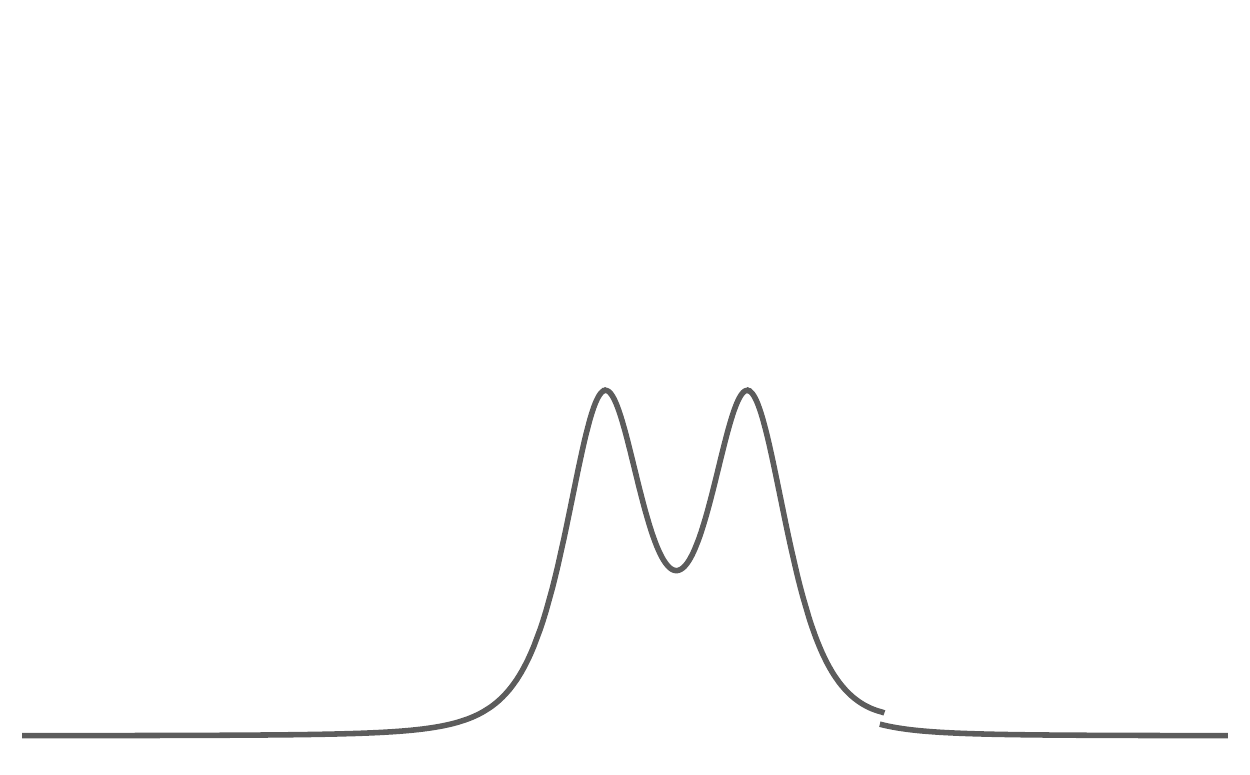}
\put(0,56){$t=t_1>0$}
\put(59.5,28){\vector(1,0){9}}
\put(49,28){\vector(-1,0){9}}
\thicklines
\put(70.8,2.5){\line(0,1){63}}
\put(0,2.5){\line(1,0){103}}
\end{overpic}
\end{subfigure}
\begin{subfigure}{0.31\textwidth}
\begin{overpic}[scale=0.33]{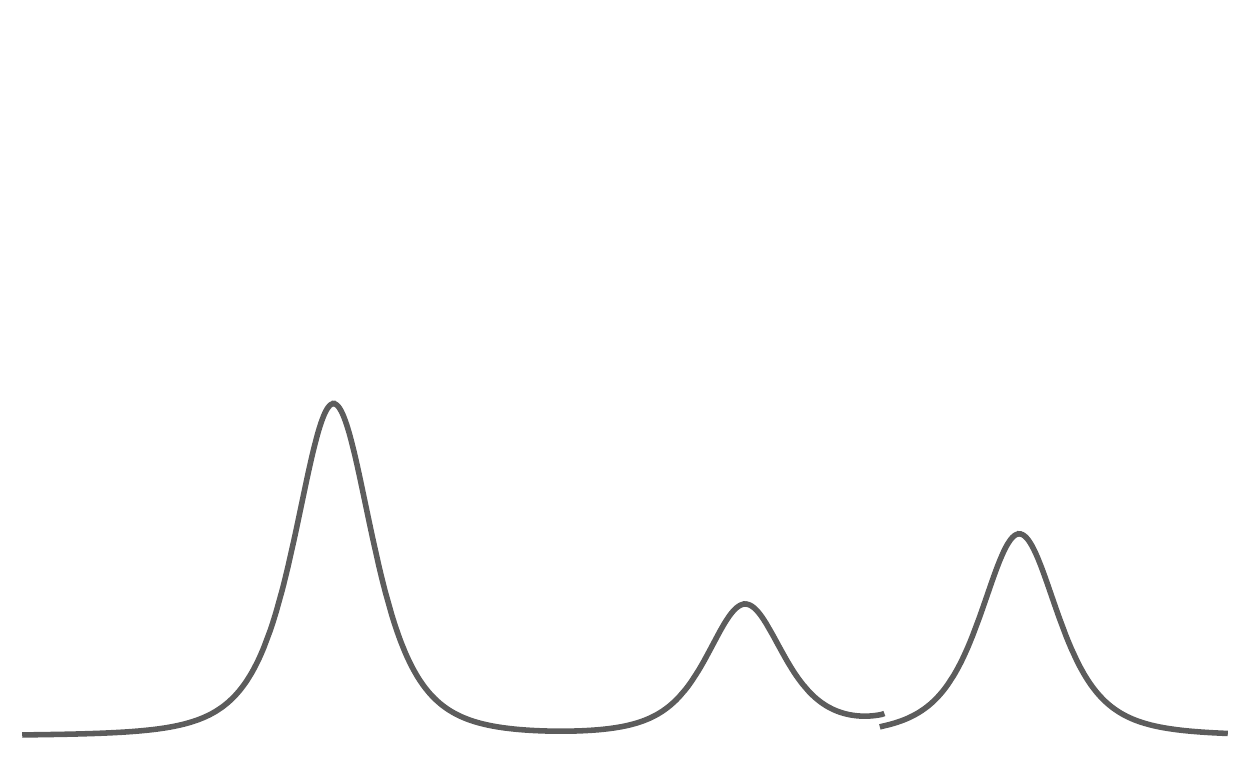}
\put(0,56){$t=t_2>t_1$}
\put(59.6,12){\vector(-1,0){9}}
\put(27,28){\vector(-1,0){9}}
\put(81,17.5){\vector(1,0){9}}
\thicklines
\put(70.8,2.5){\line(0,1){63}}
\put(0,2.5){\line(1,0){103}}
\end{overpic}
\end{subfigure}
\caption{Three snapshots in the time evolution of a state created by acting with a local operator on the vacuum at $t=0$, at some position on the left of the interface, with a finite shift in imaginary time. The curves show the expectation value of the energy density.  The pictures are obtained using a two-particle state in the free boson example of subsection \ref{subsec:boson}. 
}
\label{fig:frames}
\end{figure}
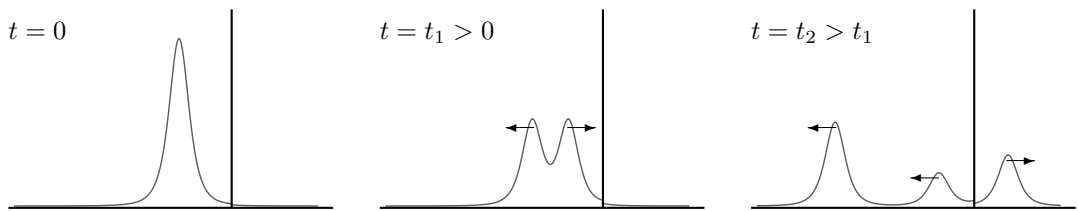

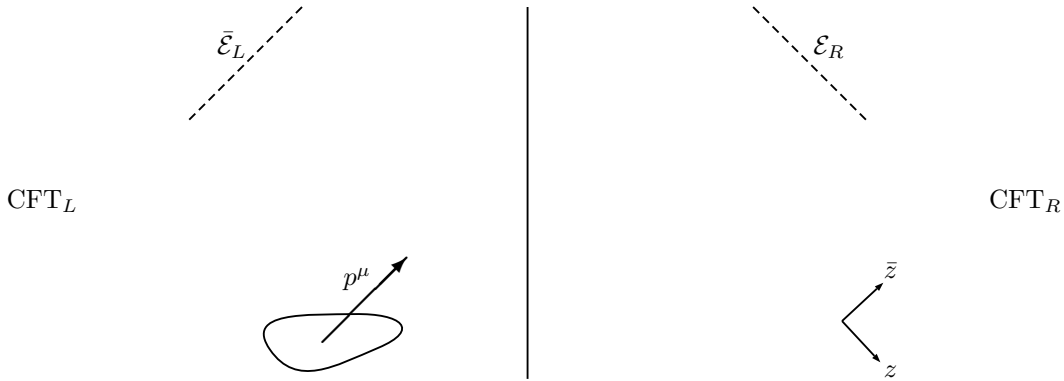
\begin{figure}[t]
\centering
\begin{overpic}[width=0.6\textwidth]{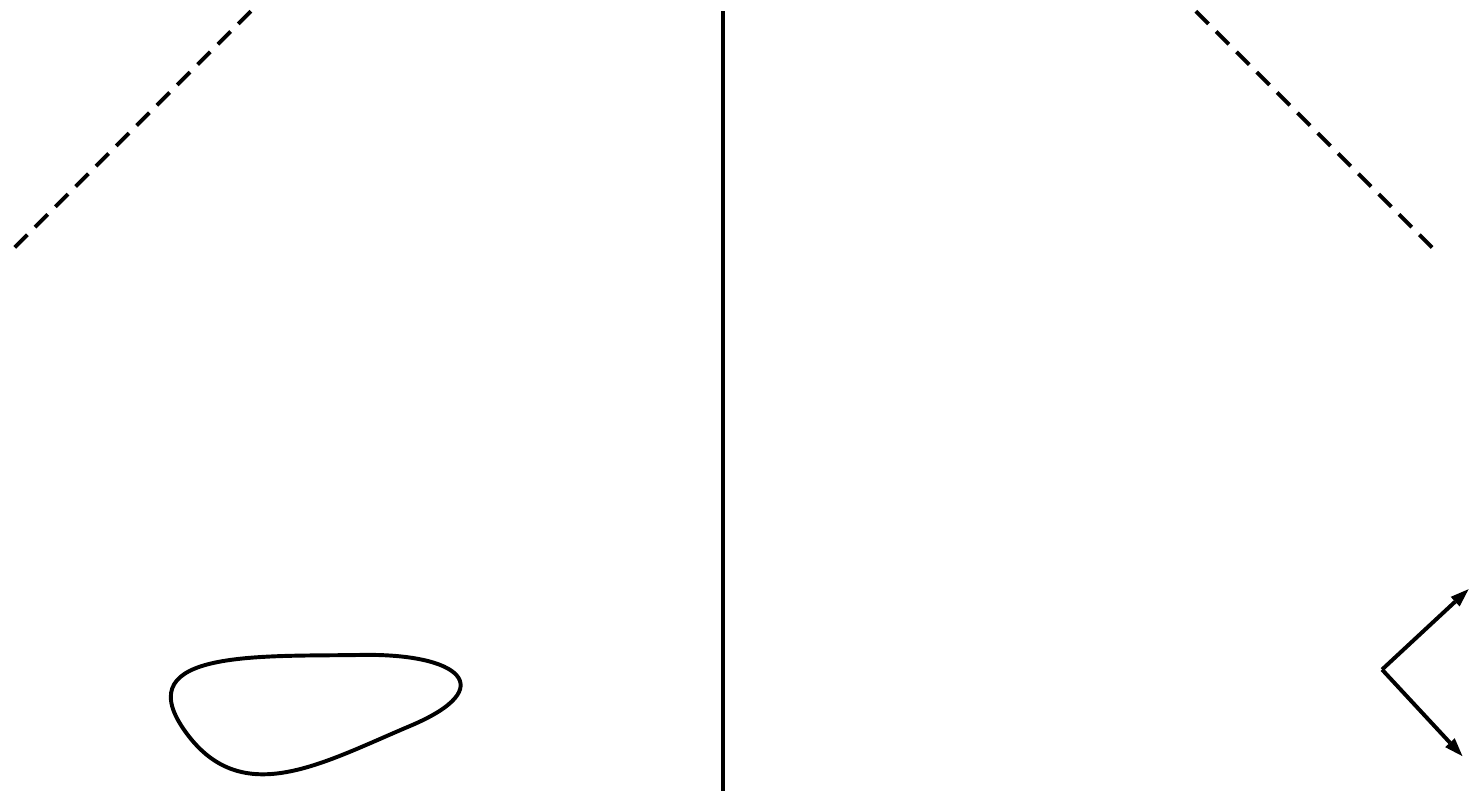}
\put(-25,25){CFT$_L$}
\put(115,25){CFT$_R$}
\put(5,47){$\bEop_L$}
\put(90,47){$\Eop_R$}
\put(100,1){$z$}
\put(100,15){$\bz$}
\put(23,14){$p^\mu$}
\thicklines
\put(20,6){\vector(1,1){12}}
\end{overpic}
\caption{The scattering experiment which defines the reflection and transmission coefficients. The expectation values of the ANEC operators $\bEop_L$ and $\Eop_R$ are computed in a state created by acting with a local operator on a compact region in the far past. We may increase the size $\ell$ of the wave function, as long as $\ell<D$, and create a state with an approximate momentum $p^\mu$.}
\label{fig:scatteringSetup}
\end{figure}

Let us prepare a state by acting with a local operator as in eq. \eqref{stateO} on the left of the interface, at some time $t=0$. As we follow the time evolution, the peak splits into a left  and a right moving part. While the former propagates away, the latter hits the interface, and splits again in a reflected and a transmitted wave -- see fig. \ref{fig:frames}.
We would like to measure the amount of energy and charge contained in these waves. However, if we choose a generic wave function, the state is significantly affected by the interface already at $t=0$. Indeed, the energy density has power-law tails and does not vanish at the location of the interface and on its right. Therefore, we need a way to isolate the reflection and transmission processes. Ideally, we would like to prepare the state very far away from the interface. We shall implement this via a limiting procedure.
We choose a normalized wave function with compact support of size $\ell$:
\begin{equation}
\int_{-\infty}^\infty\! |f(x)|^2\, dx=1~, \qquad  f(x)=0\ \ \textup{if} \ \ |x|>\ell~.
\label{wavefunctionf}
\end{equation}
Then we fix ideas by placing the perturbation on the left, so we pick a local operator $O_L$ belonging to the CFT$_L$. Let us stress that, unless stated otherwise, $O_L$ is an unrestricted linear combination of quasi-primaries and descendants, \emph{i.e.} it creates a generic state in the Hilbert space of the CFT$_L$. Obviously, an analogous construction exists with matter incoming from the right. We then choose a second scale, $D>\ell$, and define a one-parameter family of states:
\beq
\keti{O_L,D}= \int \! d^2x \, f(z)
f(\bz+D) O_L(z,\bz)\keti{0}~.
\label{wavepacket}
\eeq
Here and in the rest of the paper, the subscript I on a state means that it belongs to the constant time Hilbert space of the defect CFT, which contains the impurity at $x=0$. Similarly, the same subscript denotes the correlation functions computed in the presence of the defect. While the state \eqref{wavepacket} depends on the specific choice of $f$, the reflection and transmission coefficients will not, therefore we only denoted explicitly the important scale $D$. In fact, one may even apply the operator at a single point in spacetime, with a finite shift in imaginary time. As we increase $D$, the operator $O_L$ is applied in the far past in a region further and further away from the interface, so we gain control over the state of the system in the past.\footnote{The reason for choosing a frame where the operator is sent to the past along a light ray, rather than just far away at constant time, is technical: we want the scattering of the wave packet against the interface to happen around some fixed position along the $z$ axis, independent of the value of $D$.} The setup, depicted in fig. \ref{fig:scatteringSetup}, clearly resembles a scattering experiment. 
Let us stress that we do not need the limiting state $\keti{O_L,\infty}$ to be part of the Hilbert space. The $D\to \infty$ limit will always be understood in the weak sense: as a limit of overlaps and matrix elements, rather than of states.
For instance, we can compute the limit of the norm and obtain:
\beq
\lim_{D\to\infty} \braketi{O_L,D|O_L,D}=\braket{O_L,D|O_L,D}~,
\label{norm}
\eeq
where $\ket{O_L,D}$ is the state obtained as in eq. \eqref{wavepacket}, but now acting on the translational invariant vacuum of the CFT$_L$, \emph{i.e.} in the absence of the interface. Its norm is in fact independent of $D$. It is simple to derive eq. \eqref{norm}. The norm is determined by a two-point function, evaluated in a region where the $O_L\times O_L$ OPE converges. The identity contribution is the only one left in the strict $D=\infty$ limit, so the norm of the state can be computed in the translational invariant theory.

\begin{figure}[]
\centering
\begin{tikzpicture}[scale=1.1]
\draw [thick] (0,-2) -- (2,0);
\draw  [thick] (2,0) -- (0,2) ;
\draw [thick] (0,2) -- (-2,0);
\draw [thick] (-2,0) -- (0,-2) ;
\draw [thick] (0,2) -- (0,-2);
\draw [dashed] (-2,0) -- (2,0);
\draw [red,decorate,decoration={snake,amplitude=0.8pt}] (-1.3,-0.7) -- (0.7,1.3);
\draw [decorate,decoration={snake,amplitude=0.8pt}] (-0.7,-1.3) -- (0,-0.6);
\draw [decorate,decoration={snake,amplitude=0.8pt}] (0.7,-1.3) -- (-1.3,0.7);
\draw [red,decorate,decoration={snake,amplitude=0.8pt}] (0,0.6) -- (-0.7,1.3);
\node at (0.7,-1.3) [below] {$1$};
\node at (-0.7,-1.3) [below] {$2$};
\node at (-1.3,-0.7) [below] {$3$};
\node at (-1.3,0.7) [above] {$4$};
\node at (-0.7,1.3) [above] {$5$};
\node at (0.7,1.3) [above] {$6$};
\draw [thick,->] (3,-1) -- (3.6,-1.6) node[anchor=west] {$w$};
\draw [thick,->] (3,-1) -- (3.6,-0.4) node[anchor=west] {$\bar{w}$};
\node at (-0.6,0) [draw,diamond,fill=orange!50] {}; 
\end{tikzpicture}
\caption{The Penrose diagram associated to the time evolution of $\keti{O_L,D}$, for some finite $D$, up to a time translation which makes explicit the time reversal invariance of the process. In the picture the operator is applied at $t=0$. The wavy lines denote the the bumps in the expectation value of the energy density, and the orange diamond denotes the support of the wave packet. The numbering of the lines refers to table \ref{tab:singularities}.}
\label{fig:penroseStory}
\end{figure}
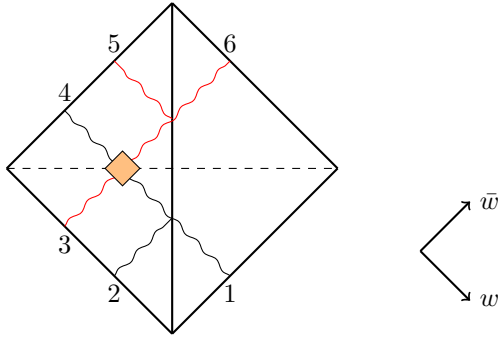

Before moving to the definition of the reflection and transmission coefficients, we would like to look back at the time evolution of the state \eqref{wavepacket}. It is easy to see that the full history of the state is as depicted in fig. \ref{fig:penroseStory}.
The red wavy lines are the bumps we are interested in, and the role of the $D\to\infty$ limit is to separate them from the other regions of non vanishing energy density. It is interesting to notice that a scattering process has happened in the past, in order to create the simple situation depicted in the $t=0$ frame in fig. \ref{fig:frames}. The left and right moving black bumps which scatter on the interface are perfectly entangled so that the result is the unique left moving bump which we see at $t=0$. The process can be easily computed explicitly in the free examples of section \ref{sec:examples}, but we can understand it in general as the result of time reversing the reflection and transmission process which we study in this work. 

The bumps in the energy density correspond to singularities in the three-point function $\braketi{O_LT_{\mu\nu}O_L}$, ordered as written. Some of these singularities are determined by the OPE, while others are singularities which result from light rays being reflected or transmitted by the interface \cite{Maldacena:2015iua,Lauria:2017wav}. We summarize the situation in table \ref{tab:singularities}.
\begin{table}[h]
\centering
\begin{tabular}{clcc}
\toprule
       & position      &    bump      &       \\
       \midrule
$\bT_R(\bz)$ & $\bz=\bz_i$  & 1 &  \\     
 \hline
\multirow{2}*{$T_L(z)$} & $z=-\bz_i$   & 2 &  \\   
                & $z=z_i$   & 3 &  OPE  \\  
 \hline               
\multirow{2}*{$\bT_L(\bz)$} & $\bz=\bz_i$   & 4 &  OPE \\   
                       & $\bz=-z_i$   & 5 &   \\   
                        \hline
$T_R(z)$ & $z=z_i$   & 6 &  \\                                 
       \bottomrule
\end{tabular}
\caption{Relevant singularities in the three-point function $\braketi{O_L(z_1,\bz_1)T_{\mu\nu}(z,\bz)O_L(z_2,\bz_2)}$, for fixed $(z_i,\bz_i)$ with $i=1,2$. For each component of the stress tensor, we report the position of the singularities, the bump they cause as numbered in fig. \ref{fig:penroseStory}, and their nature.}
 \label{tab:singularities}
\end{table}
The singularities which are not governed by an OPE are in general branch points, as we shall see in the example of subsection \ref{subsec:ising}. What makes the reflection and transmission processes simple and universal is that even these singularities become poles when we take the limit $D\to\infty$. We shall prove this statement in subsection \ref{subsec:nonholo}, but first let us define our observables.

\subsection{The observable}
\label{subsec:observable}

Having described the state where we want to perform the measurement, it is easy to define the energy reflection and transmission coefficients:
\begin{equation}
\begin{split}
&\ourT_L = \lim_{D\to\infty}\frac{\braketi{O_L,D|\Eop_R|O_L,D}}{\braket{O_L,D|\Eop_L|O_L,D}}~, \\
&\ourR_L =\lim_{D\to\infty}\frac{\braketi{O_L,D|\bEop_L|O_L,D}-\braket{O_L,D|\bEop_L|O_L,D}}{\braket{O_L,D|\Eop_L|O_L,D}}~, \\
&\ourT_R = \lim_{D\to\infty}\frac{\braketi{O_R,D|\bEop_L|O_R,D}}{\braket{O_R,D|\bEop_R|O_R,D}}~, \\
&\ourR_R =\lim_{D\to\infty}\frac{\braketi{O_R,D|\Eop_R|O_R,D}-\braket{O_R,D|\Eop_R|O_R,D}}{\braket{O_R,D|\bEop_R|O_R,D}}~.
\end{split}
\label{TRenergyhhb}
\end{equation}
Let us point out a few features of this definition. 
Firstly, $\ourT_R$ and $\ourR_R$ measure reflection and transmission in a state created by an operator of the CFT$_R$ with a construction analogous to eq. \eqref{wavepacket}.
Secondly, the states $\ket{O_L,D}$ and $\ket{O_R,D}$ are created in the same way as their counterparts with the subscript $\rm{I}$, but this time on top of the translational invariant vacuum of the left or right CFTs. This means in particular that the flux of energy in these states is independent of $D$.  Finally, recall that the subscript $L$ or $R$ of the ANEC operator means that the corresponding stress tensor is employed in the definition \eqref{energyOperators} -- see fig. \ref{fig:penroseDefect}. The meaning of the definition is obvious given the considerations in section \ref{sec:propagation} and subsection \ref{subsec:wflim}: the flux of energy at infinity is normalized by the energy of the red bump in fig. \ref{fig:penroseStory}, which travels towards the interface in the far past. As we send $D\to\infty$, this incoming energy can be computed in the absence of the interface. Additionally, we subtract from the reflection coefficient the fraction of energy which flows away, carried by the black bump in fig. \ref{fig:penroseStory}, again estimated without the interface. Of course, all states must be normalized, but the norms all cancel out due to eq. \eqref{norm}.

While this definition makes intuitive sense, it should pass at least two basic tests. First, reflection and transmission should sum up to one, since, as we mentioned, the interface lacks a local stress tensor, and therefore cannot absorb energy. This is a check that the $D\to\infty$ limit correctly disentangles the relevant process. Second, the transmission coefficients should vanish for a factorizing interface and should equal one for a topological one:
\begin{equation}
\ourT+\ourR=1~, \notag
\end{equation}
\begin{equation}
\left\{
\begin{array}{ll}
\textup{factorizing interface } &\implies \ourT_L=\ourT_R=0~,  \\
\textup{topological interface} &\implies \ourT_L=\ourT_R=1~. 
\end{array}
\right.
\label{wishlist}
\end{equation}
The items in the wish list are true thanks to the gluing conditions \eqref{gluingT}.
In order to fix ideas, let us focus on the coefficients $\ourT_L,\,\ourR_L$, \emph{i.e.} we create the state with a local operator belonging to CFT$_L$. Let us adopt the following definition, to avoid cluttering:
\beq
G_{L/R}(z)=\braketi{O_L(z_1,\bz_1) T_{L/R}(z) O_L(z_2,\bz_2)}~,
\eeq
and similarly for $\bar{G}_{L/R}$ with $\bar{T}_{L/R}$. Then, eq. \eqref{gluingT} implies the following:
\beq
\bar{G}_L(\bz)+G_R(-\bz)=\sum_{i=1}^2\,\textup{Res}_{w=z_i}\!\left[\frac{G_L(w)}{-w-\bz}\right] -\sum_{i=1}^2\,\textup{Res}_{\bar{w}=\bz_i}\!\left[\frac{\bar{G}_L(\bar{w})}{\bar{w}-\bz}\right] ~.
\label{barGLGRup}
\eeq
This is derived in appendix \ref{app:OOT} for the Euclidean correlator -- see eqs. \eqref{GLGR} and \eqref{barGLGR} -- but the relation remains valid after the analytic continuations required to reach the relevant Lorentzian kinematics. Eq. \eqref{barGLGRup} encodes a new useful information about the singularities described in table \ref{tab:singularities}. In the sum $\bT_L(\bz)+T_R(-\bz)$, the singularity at $\bz=-z_1(z_2)$ is a pole determined by the OPE $T_L\times O_L$, despite the individual singularities are not of the OPE type. This is the technical way in which conservation of energy in the reflection and transmission process is encoded in the correlator. Starting from eq. \eqref{barGLGRup}, it is easy to derive $\ourT_L+\ourR_L=1$, by just performing the integrals which define $\bEop_L$ and $\Eop_R$. The contours can be closed on either side, because the pole at infinity vanishes exactly by time translational invariance. 
Then the sum $\ourT_L+\ourR_L$ is determined by the simple poles on the r.h.s. of eq. \eqref{barGLGRup}. Crucially, the residues are fixed by translational invariance in terms of derivatives of the lower point function $\braketi{O_L(x_1)O_L(x_2)}$. Here, again, the $O_L\times O_L$ OPE can be used to conclude that in the $D\to \infty$ limit the interface disappears, and the result follows. Notice that the computation only relies on the gluing conditions and on the simple pole in the OPE of the stress tensor with a local operator, and is therefore valid in any state of the kind \eqref{wavepacket}.
An identical computation proves the same for a state created on the right. 

From now on, therefore, we shall focus on the transmission coefficient.
It is then trivial to show that also the second point in the wishlist \eqref{wishlist} is met: $\ourT=0$ for a factorizing interface and 1 for a topological one. The first result is a consequence of the following simple equation:
\beq
G_R(z)=0~, \quad \textup{factorizing interface,}
\eeq 
which is derived in appendix \ref{app:OOT}, eq. \eqref{GGGGbound}. In fact, the right stress tensor is not correlated with any  of the left local operators if the interface is factorizing. Analogously, we derive that $\ourT=1$ for a topological interface from the following -- see eq. \eqref{GGGGtopo}:
\beq
G_L(z) = G_R(z) = -\Sigma_i\,\textup{Res}_{w=z_i}\!\left[\frac{G_L(w)}{w-z}\right] ~, \quad \textup{topological interface.}
\eeq 
The stress tensor is insensitive to a topological interface. Let us anticipate an important comment. While we have proven the implications in eq. \eqref{wishlist}, the reverse implications are true as well.  We shall prove this statement in subsection \ref{subsec:bounds}.

The definition \eqref{TRenergyhhb} is ready to be used, but it requires performing multiple integrals and limits on a three point function in the presence of the interface. This is in general a complicated object, a function of two cross-ratios. In the following, we shall show that, remarkably, $\ourT$ and $\ourR$ do not depend on the wave function used to construct the state \eqref{wavepacket}, and are determined by a single piece of CFT data, which has been previously studied in \cite{Quella:2006de}. The final result is contained in eq. \eqref{TLsl2general}, but let us first describe a special case.

\subsection{Energy reflection in a holomorphic state}
\label{subsec:holo}

Let us consider a state created, as in eq. \eqref{wavepacket}, with a holomorphic operator $O_L(z)$ belonging to the CFT$_L$. In particular, $O_L$ might be the stress tensor $T_L$, which always exists. As we discussed in section \ref{sec:propagation}, a holomorphic operator does not excite left movers when acting on the translational invariant vacuum. Therefore, the black bump in fig. \ref{fig:penroseStory} is absent. Accordingly, the state is independent from the position of the wave packet along $\bz$, and the $D\to\infty$ limit in eq. \eqref{TRenergyhhb} can be dropped. Let us now first assume that $O_L$ is a linear combination of quasi-primaries. Then the reflection and transmission coefficients are determined by the following three-point function
\beq
\braketi{O^1_L(z_1)T_R(z)O^2_L(z_2)}=
 \frac{c_{12T_R}}{(z_1-z_2)^{h_1+h_2-2}(z_1-z)^{h_1-h_2+2}(z_2-z)^{h_2-h_1+2}}~.
 \label{OOTholo}
\eeq
Here $O^1_L$ and $O^2_L$ are quasi-primaries of weights $(h_1,0)$ and $(h_2,0)$ respectively. Even in the presence of the interface, the correlator \eqref{OOTholo} is fixed up to a single number. 
The coefficient $c_{12T_R}$ can then be fixed in terms of lower point functions, by using the fusion $O^1_L \times O^2_L$. We obtain a sum over two-point functions of holomorphic operators on opposite sides of the interface. As explained in appendix \ref{app:kinematics}, only holomorphic quasi-primaries with equal weight correlate:
\beq
\braketi{O_L(z_1)O_R(z_2)}=\frac{b_{O_LO_R}\, \delta_{h_L,h_R}}{(z_1-z_2)^{2h_L}}~.
\label{OLORholo}
\eeq
Therefore, $c_{12T_R}$ is fixed in terms of the correlators of quasi-primaries of weight $(2,0)$ in the CFT$_L$ with the stress-tensor $T_R$. Let us now assume that there is a unique $(2,0)$ quasi-primary which is a singlet under all the symmetries preserved by the interface, \emph{i.e.} the stress-tensor $T_L$. In subsection \ref{subsec:discussion}, and in the example \ref{subsec:cosets}, we shall comment on the case where this assumption does not hold. Let us further denote with $c_{LR}$ the coefficient appearing in the two-point function
\beq
\braketi{T_L(z_1)T_R(z_2)}=\frac{c_{LR}/2}{(z_1-z_2)^{4}}~.
\label{CLR}
\eeq
We then easily find the following:
\beq
\langle O^1_L(z_1)\left(T_R(z)-\frac{c_{LR}}{c_L} T_L(z)\right)O^2_L(z_2)\rangle_\textup{I}=0~,
\label{HoloStep}
\eeq
where $c_L$ is the central charge of the CFT$_L$.  In deriving this equation we made use of the fact that, since the correlator of holomorphic operators does not depend on the distance from the interface, then
\beq
\braketi{O^1_L(z_1)T_L(z)O^2_L(z_2)}=
\braket{O^1_L(z_1)T_L(z)O^2_L(z_2)}~.
\eeq
Clearly, taking derivatives in $z_1$ and $z_2$ does not change the validity of eq. \eqref{HoloStep}, which is therefore true for generic local operators $O_L^1$ and $O_L^2$.  Therefore, by comparison with the definition \eqref{TRenergyhhb}, we obtain
\beq
\ourT_L= \frac{c_{LR}}{c_L}~.
\label{TLholo}
\eeq
Remarkably, the coefficient does not depend on the holomorphic local operator used to construct the state, nor on the choice of wave packet.  

A completely analogous computation fixes the transmission coefficient for a state created by an antiholomorphic operator belonging to the CFT$_R$:
\beq
\ourT_R= \frac{c_{LR}}{c_R}~.
\label{TRantiholo}
\eeq
In deriving this last equation, we assumed that $c_L= \bar{c}_L$ and $c_R=\bar{c}_R$. This implies in particular that $\bar{c}_{LR}=c_{LR}$, where the former is the coefficient in the two-point function $\braketi{\bT_L\bT_R}$. Relaxing the constraint is trivial, and we shall mention the result in the concluding remarks, section \ref{sec:summary}.

Let us conclude this subsection with a few comments on the intermediate eq. \eqref{HoloStep}. This equation is in fact more general than the result \eqref{TLholo}. Indeed, if we evaluate the linear combination $T_R-\ourT_L T_L$ on the interface, we obtain a local defect operator. Eq. \eqref{HoloStep} then affirms that this local operator vanishes in the subspace of the Hilbert space generated by holomorphic operators.
It is not surprising that such a defect operator does not vanish in the full Hilbert space, since the latter contains states in which matter is incoming towards the interface both from the left and from the right. One might ask if the more general combination $T_R-\ourT_L T_L-\ourR_R \bT_R$ instead vanishes in the larger subspace spanned by (products of) holomorphic operators of the CFT$_L$ and antiholomorphic operators of the CFT$_R$.\footnote{Notice that $T_R-\ourT_L T_L-\ourR_R \bT_R$ cannot vanish in a generic state. Indeed, when evaluated on the defect, this combination is a local operator whose two-point function does not vanish. } The answer is again negative. In the generic state of this subspace, right and left movers are entangled. Interference effects are then responsible for the failure of the putative eq. $T_R-\ourT_L T_L-\ourR_R \bT_R=0$, as it is easy to verify in the free theory examples of section \ref{sec:examples}. In fact, the same interference effects are at play in the history of a generic state as described in fig. \ref{fig:penroseStory}. If we look at the black bump, we see that the right and left movers labeled $1$ and $2$ do not scatter independently with the interface: there is no energy being transmitted towards the right.

Let us now go back to the reflection and transmission coefficients as defined in eq. \eqref{TRenergyhhb}, and tackle the most general case.

\subsection{Energy reflection in a generic state}
\label{subsec:nonholo}

In this subsection, we generalize the analysis to states created by generic local operators. We shall show that eqs. \eqref{TLholo} and \eqref{TRantiholo} are still valid in this general case.

Let us start again with a state created by a linear combination of quasi-primary operators $O^i_L$ of dimensions $(h_i,\bh_i)$, belonging to the CFT$_L$. The correlation function of two quasi-primaries with the stress tensor $T_R$ contains two cross ratios:
\begin{multline}
\braketi{O^1_L(z_1,\bz_1)T_R(z)O^2_L(z_2,\bz_2)}= \\
\left(\frac{(z_2+\bz_2)(z_1-z)}{(z_1+\bz_1)(z_2-z)}\right)^{\bh_1-\bh_2}\!\!\!\!\!\!
 \frac{f(\xi,u)}{(z_1-z)^{h_1-h_2+2}(z_2-z)^{h_2-h_1+2}(z_1-z_2)^{h_1+h_2-2}(\bz_1-\bz_2)^{\bh_1+\bh_2}}~,\\
 \xi=\frac{z_{12}z_{\bar{1}\bar{2}}}{(z_1+\bz_1)(z_2+\bz_2)}~,\ 
 u=\frac{z_{12}(z+\bz_1)}{(z_1+\bz_1)(z-z_2)}~,
 \label{OOTR}
\end{multline}
Like in the previous subsection, the strategy will involve the use of the $O_L^1\times O_L^2$ OPE, but this time we need to be more careful with the analytic properties of the correlator and the region of convergence of the OPE.
Let us first make the $\ii \ep$ prescription explicit:
\begin{align}
&z_1 \to z_1+\ii \ep~,  \qquad \bz_1 \to \bz_1-\ii \ep~, \notag\\
&z_2 \to z_2-\ii \ep~, \qquad  \bz_2 \to \bz_2+\ii \ep~, \label{iepsilon}
\end{align}
while we keep $z \in \mathbb{R}$. This ensures that the correlator is evaluated with the ordering shown in eq. \eqref{OOTR}, and that the states created by the operator $O$ acting on the left and right vacua are normalizable. As long as 
$\ep$ is finite, the correlation function is analytic for $\left| \Im z \right| <\epsilon $ 
\cite{Hartman:2015lfa}. 
We assume that the correlator stays analytic in $z$ also in the $D\to \infty$ limit, which corresponds to $\bz_1,\bz_2\to - \infty$. This is true in examples, and we believe it to be a mild assumption, given the physical picture developed in subsection \ref{subsec:wflim}.
In the whole derivation, we shall keep $\ep$ finite. While the state $\keti{O_L,D}$ now depends on $\ep$, the transmission coefficient will not.

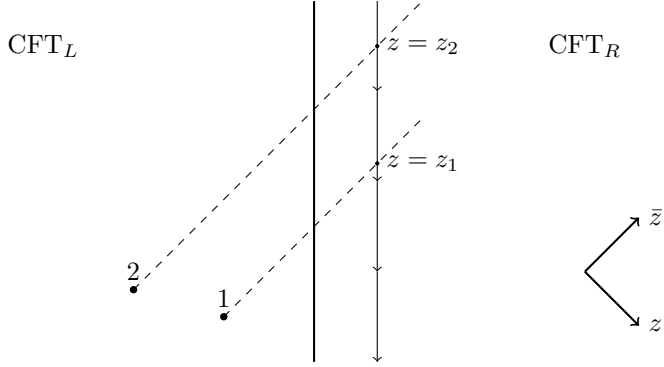
\begin{figure}[]
\centering
\begin{tikzpicture}[scale=1.2]
\draw [thick] (0,-2) -- (0,2) ;
\filldraw [black] (-1,-1.5) circle [radius=1pt] node[above] {$1$};
\filldraw [black] (-2,-1.2) circle [radius=1pt] node[above] {$2$};
\draw  [dashed] (-1,-1.5) -- (1.2,0.7);
\draw  [dashed] (-2,-1.2) -- (1.2,2);
\draw[<-] (0.7,-2) -- (0.7,-1) ;
\draw[<-] (0.7,-1) -- (0.7,0) ;
\draw[<-] (0.7,0) -- (0.7,1) ;
\draw[<-] (0.7,1) -- (0.7,2) ;
\filldraw [black] (0.7,0.2) circle [radius=0.5pt] node[right] {$z=z_1$};
\filldraw [black] (0.7,1.5) circle [radius=0.5pt] node[right] {$z=z_2$};
\draw [thick,->] (3,-1) -- (3.6,-1.6) node[anchor=west] {$z$};
\draw [thick,->] (3,-1) -- (3.6,-0.4) node[anchor=west] {$\bar{z}$};
\node at (-3,1.5) {CFT$_L$};
\node at (3,1.5) {CFT$_R$};
\end{tikzpicture}
\caption{The arrows mark the path of the operator $T_R(z)$, at fixed $z_1,\,\bz_1,\,z_2,\,\bz_2$. Due to holomorphy, the path can be freely deformed in the $\bz$ direction. The marked points along the path, at $z=z_1$ and $z=z_2$, are the Landau singularities listed in table \ref{tab:singularities}.}
\label{fig:spacetimecontour}
\end{figure}

\begin{figure}
\centering
\begin{tikzpicture}[scale=1.4]
\fill [orange!50] (-3,-2) rectangle (0,2); 
\draw [decorate,decoration={snake,amplitude=0.8pt}] (0,0) -- (2,0);
\filldraw [black] (-1,0) circle [radius=1pt] node[left] {$0$};
\filldraw [black] (0,0) circle [radius=1pt] node[left] {$1$};
\draw [dashed,postaction = {decoration={markings, mark=between positions 0.2 and 0.9 step 1cm  with {\arrow{>};}},decorate}] (-1,0) arc [start angle=160, end angle=380, radius=1.064];
\draw [postaction = {decoration={markings, mark=between positions 0.2 and 0.9 step 1cm with {\arrow{<};}},decorate}] (-1,0) arc [start angle=160, end angle=20, radius=1.064];
\draw [black] (1.9,1.8)-- (1.7,1.8) -- (1.7,2);
\node at (1.8,1.9) {$u$};
\end{tikzpicture}
\caption{The path in cross ratio space corresponding to the following configuration in the correlator \eqref{OOTR}: $\bz_{1,2}=-\infty$, $\bz_{12}$ finite, $z\in (-\infty,+\infty)$ with the $\ii \ep$ prescription of eq. \eqref{iepsilon}. In this case, $\xi=0$, and the path is a circle with radius $1+(z_1-z_2)^2/4 \ep^2$. The region $\Re u<1$, shaded in orange, is the region of convergence of the $O_L\times O_L$ OPE. As usual, the $\ii \ep$ shift prescribes how to avoid the singular points, located at $u=1$ and $u=\infty$. In general, $u=1$ and $u=\infty$ are branch points, so that the path starts on the second sheet -- the dashed part -- and ends in the Euclidean region -- the solid part. Compare with the path in spacetime, fig. \ref{fig:spacetimecontour}. The picture is qualitatively the same at finite $\xi$, but at $\xi=0$ the branch points reduce to poles.}
\label{fig:crosscontour}
\end{figure}
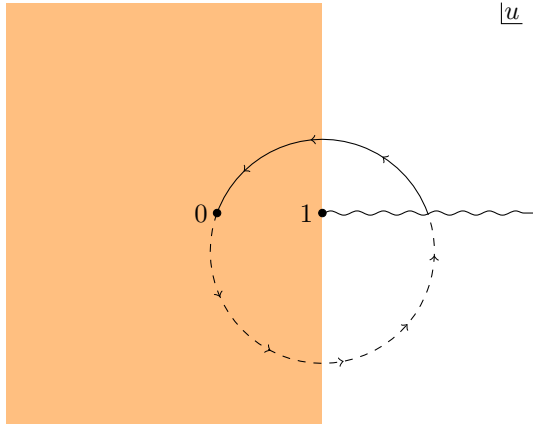

The ANEC operator $\Eop_R$ is defined by integrating $z$ on the real axis. At fixed $z_1,\,z_2,\,\bz_1,\,\bz_2$, one possible contour in spacetime is shown in fig. \ref{fig:spacetimecontour}.
It is important that part of the contour lies in a region where $T_R$ is spacelike separated from $O^1_L$ and $O^2_L$. Therefore, the $O^1_L\times O^2_L$ OPE converges here. The contour in cross ratio space is a circle in the $u$ plane at fixed $\xi$, since $u$ is an $SL(2,\mathbb{C})$ transformation of $z$ -- see eq. \eqref{OOTR}. We draw the contour in fig. \ref{fig:crosscontour} for the limiting value $\xi=0$, which corresponds to $\bz_1,\,\bz_2=-\infty$ with $\bz_{12}$ finite, together with the region of convergence of the $O^1_L\times O^2_L$ OPE. The latter is discussed in appendix \ref{app:OOTOPE}. 
The special points $u=1$ and $u=\infty$ correspond to the singularities $z=z_1$ and $z=z_2$. While in the absence of the interface these are poles, they are generically branch points in a defect CFT.
A crucial simplification occurs at $\xi=0$, \emph{i.e.} in the $D\to\infty$ limit. 

Consider first the region where the $O^1_L\times O^2_L$ OPE converges. This is best characterized using the following $\r$ coordinates:
\beq
\xi = \frac{4 \rho \rhob}{(1-\rho \rhob)^2}~, \qquad 
u=\frac{2\rho(1-\rhob)}{(1+\rho)(1-\rho \rhob)}~.   \label{xitorho}
\eeq
As we explain in appendix \ref{app:OOTOPE}, the $O^1_L\times O^2_L$ OPE converges in the region $\r,\,\rb \in \mathbb{C}$, $|\r|<1$, $|\rb|<1$.\footnote{In fact, the region described by these bounds is optimal only at $\xi=0$. For finite $\xi$, the $O_L\times O_L$ OPE converges in a strictly larger region. See appendix \ref{app:OOTOPE} for details.} It has the familiar structure
\beq
O^1_L(\rho,\rhob)\,O^2_L(0,0) \sim \rho^{-h_1-h_2+h_\textup{exch}}\rhob^{-\bh_1-\bh_2+\bar{h}_\textup{exch}} O_\textup{exch}(0,0)+\dots
\eeq
As explained in eqs. (\ref{OOholo},\ref{b12constraint}), only quasi-primaries with spin 2 correlate with the stress tensor of the CFT$_R$, therefore only the corresponding conformal blocks contribute to the function $f(\xi,u)$ in eq. \eqref{OOTR}. Now, the $\xi\to 0$ limit, at fixed $u$, corresponds to the lightcone  limit $\rhob\to0$  at fixed $\rho$. In this limit, only the blocks with $\bh_{exch}=0$ survive. We are left with the set of quasi-primaries with weights $(h,\bh)=(2,0)$, and for the moment we assume that this set consists of the stress tensor $T_L$ only. The contribution of the conformal block of the stress tensor is easily computed by noticing that it has to coincide with the whole correlator in the case of a trivial defect. Therefore:
\begin{multline}
\lim_{D\to\infty} \braketi{O^1_L(z_1+\ii \ep,\bz_1-D-\ii \ep)T_R(z)O^2_L(z_2-\ii \ep,\bz_2-D+\ii \ep)} \\
=
\frac{c_{LR}}{c_L} \braket{O^1_L(z_1+\ii \ep,\bz_1-\ii \ep)T_L(z)O^2_L(z_2-\ii \ep,\bz_2+\ii \ep)}~.
\label{OOTxizero}
\end{multline}
The coefficient $c_{LR}$ was defined in eq. \eqref{CLR}. Recall that the absence of the subscript I in the r.h.s. means that the correlator is computed in the translational invariant vacuum of the CFT$_L$. In particular, the r.h.s. vanishes if $O_L^1 \neq O_L^2$.
While this equation was derived in the region of convergence of the $O^1_L\times O^2_L$ OPE, the correlator is analytic in $u$ in a neighborhood of the integration contour, and part of the contour runs in the region where \eqref{OOTxizero} holds. We conclude that eq. \eqref{OOTxizero} is valid for all $z \in \mathbb{R}$. It is now not hard to convince oneself that eq. \eqref{OOTxizero} is valid even if $O_L^1$ and $O_L^2$ are not quasi-primaries.\footnote{Both $\xi$ and $u$, when evaluated with the shifts $\bz_i\to \bz_i-D$, have a regular Taylor expansion around $D=\infty$. Therefore, derivatives of $\xi$ vanish at least as fast as $\xi$. In the region of convergence of the $O_L^1\times O_L^2$ OPE, the function $f(\xi,u)$ in eq. \eqref{OOTR} starts with a constant if the fusion contains the stress tensor $T_L$, otherwise is suppressed by a power of $\xi$. Therefore, derivatives in $z_1,\, z_2,\,\bz_1 ,\,\bz_2$ are suppressed, unless $\bh_1=\bh_2$ and they are all applied to the kinematical prefactors in eq. \eqref{OOTR}. But this precisely yields the functional form of the correlator in the translational invariant theory.}

The only wrinkle left to discuss is whether we can switch the integral in $z$ with the $D\to \infty$ limit, \emph{i.e.} if the order $\xi$ corrections to eq. \eqref{OOTxizero} can change the final result.
Although we do not have enough control over such corrections, it is physically clear that the order in which we perform the two operations is irrelevant. The right hand side of eq. \eqref{OOTxizero} computes the limiting value of the energy deposited at a specific point at null infinity. By summing up the contribution at every point, we will obtain the limiting value of the total transmitted energy. The only way this may fail is if a fraction of energy which stays finite as $D\to\infty$ was spread over a region whose position or size depends on $D$: for instance, a region centered around $z \sim D$ or whose size grows with $D$. But we saw in section \ref{sec:propagation} that energy is transported rigidly along a lightlike trajectory: there is no spread. It is clear from fig. \ref{fig:penroseStory} that if we perform a set of experiments with identical wave packets which are further and further away along the $\bz$ direction, the fraction of transmitted energy will be closer and closer to the limiting case.\footnote{For comparison, consider the reflection coefficient. If we replace $T_R \to \bT_L$ in eq. \eqref{OOTR}, the correlator is singular when $z=\bz_1$ and $\bz_2$  -- see table \ref{tab:singularities}. These poles are responsible for the bump labeled 4 in fig. \ref{fig:penroseStory}, which is moving away as $D$ grows. In this case the $D\to\infty$ limit and the integral do not commute. Technically, the hypotheses of the dominated convergence theorem are violated. Notice however that the contribution of this bump is precisely the one subtracted out in the definition \eqref{TRenergyhhb}.}   We can therefore safely assume that the transmitted energy is completely determined by eq. \eqref{OOTxizero}.
By comparison of eq. \eqref{OOTxizero} with the definition \eqref{TRenergyhhb}, we conclude that
\beq
\ourT_L = \frac{c_{LR}}{c_L}~, \quad \ourT_R = \frac{c_{LR}}{c_R}~, \qquad \textup{where} \quad
\braketi{T_L(z_1)T_R(z_2)}=\frac{c_{LR}/2}{(z_1-z_2)^{4}}~,
\label{TLsl2general}
\eeq
for any state constructed as in eq. \eqref{wavepacket}. This is the main result of the paper. It shows that the fraction of energy transmitted through the interface is a constant, which only depends on the central charge and on a single piece of defect CFT data, the coefficient $c_{LR}$ of a two-point function. As a side comment, it is interesting to notice that $\ourT_L=\ourT_R$ if $c_L= \bar{c}_L=c_R=\bar{c}_R$. No discrete symmetry of the interface is necessary to establish this fact: specifically, it is true even if the defect breaks time reversal symmetry, which would also exchange right and left movers. 

As we review in section \ref{sec:examples}, the constant $c_{LR}$  can be computed every time the boundary state corresponding to the interface is known. Before moving to the examples, we dedicate a few words to the discussion of the physical meaning  and of some of the consequences of this result.

\subsection{Discussion}
\label{subsec:discussion}

The universality of the result \eqref{TLsl2general} deserves some comments.
In principle, we could have expected the reflection and transmission coefficients to depend on the wave packet $f(z)$ in eq. \eqref{wavepacket}, as well as on the operator used to create the state.

In fact, independence of the coefficients from the wave packet could be expected. Indeed, let us choose a basis of wave packets which create states with well defined incoming momentum $p^\mu=(p,\bp)$. Although we cannot use plane waves due to the presence of the interface, we can approximate a plane wave arbitrarily well by taking the support $\ell$ of the wave packet defined in eq. \eqref{wavefunctionf} to be large, specifically $\ell\to \infty$ with $\ell/D\to 0$.  Due to scale invariance, $\ourT$ and $\ourR$ can only depend on the ratio $\bp/p$. But from the discussion in section \ref{sec:propagation}, we know that $p$ is the momentum carried away from the interface towards the left. Therefore we expect $\ourT_L,\ \ourR_L$ to be independent from it, thus constant.

The complete independence of the transmission and reflection coefficients from the way we act on the vacuum to create the state is more interesting. Different local operators create conformal matter with a different momentum distribution -- see eq. \eqref{holoNorm} -- but the carriers all behave exactly in the same way when scattered against a conformal interface, at least as far as the average fraction of transmitted and reflected energy is concerned. The situation is very different from a generic interactive quantum field theory, where the strength of the coupling to the interface could depend on the particle type. It would be interesting to study the reflection and transmission coefficients along an RG flow, and analyze in detail the approach to the fixed points.

The only way to violate the universality of eq. \eqref{TLsl2general} is to allow for multiple holomorphic quasi-primary operators of spin two. It is of course straightforward to modify the discussion to accommodate this more general situation. 
  For instance, if the CFT$_L$ is the direct product of independent theories, the transmission coefficient for states belonging to each sector is determined by the coupling of the stress tensor of that sector with the stress tensor of the CFT$_R$. There is at least one more interesting class of theories where multiple $(2,0)$ quasi-primaries exist, \emph{i.e.} WZW models, which enjoy an affine Lie Algebra symmetry. The fusion of two currents generates various $(2,0)$ quasi-primaries, only one of which is the Sugawara stress tensor.\footnote{In fact, if they are not $sl(2)$ descendants, these operators are actually Virasoro primaries, since the only weight $(0,0)$ Virasoro primary is the identity. Since a primary is in particular a quasi-primary, in this paper we do not keep track of the distinction unless necessary.} Operators belonging to different representations of the affine symmetry couple differently to the additional spin two holomorphic operators. As a consequence, the reflection and transmission coefficients are now state dependent. We consider one example of this kind in subsection \ref{subsec:cosets}, but let us make now two general comments. Firstly, eq. \eqref{TLsl2general} still computes the transmission coefficients in a state created by the stress tensor, a result that is completely universal, and simply follows from the fact that the fusion of the identity with itself only contains the identity. Secondly, this class of examples fits in well with the following qualitative picture. When the CFT has a global symmetry, different states can be characterized by the additional conserved quantities. Correspondingly, conformal matter with different quantum numbers is reflected and transmitted differently across a conformal interface.

\subsection{Bounds on reflection, the relation with \cite{Quella:2006de} and the displacement operator}
\label{subsec:bounds}

We should expect the fractions of energy reflected and transmitted to be both non negative.
The simple inequalities $0\leq\ourT_{L/R}\leq 1$ can be refined in light of eq. \eqref{TLsl2general}:
\beq
0\leq \ourT_L \leq \min \left(1,\frac{c_R}{c_L}\right)~, \qquad
0\leq \ourT_R \leq \min \left(1,\frac{c_L}{c_R}\right)~.
\label{Tbounds}
\eeq
These bounds correspond to the unique equation
\beq
0<c_{LR}<\min (c_L,c_R)~.
\label{clrbounds}
\eeq 
Therefore when $\ourT_L$ saturates one of the bounds, $\ourT_R$ saturates the corresponding one. 

The relation between the transmission coefficients and the two-point function of the stress tensors allows us to further characterize the interfaces which saturate the bounds \eqref{Tbounds}. Let us first point out that, due to the gluing conditions, all the non vanishing two-point functions of the stress tensors are fixed in terms of $c_{LR}$ and the central charges \cite{Quella:2006de,Billo:2016cpy}:\footnote{The same is true of the three-point functions as well.}
\begin{subequations}
\begin{align}
\braketi{T_L T_L} &= \braketi{\bT_L \bT_L} \propto c_L/2~, \label{TlTl}\\
\braketi{T_R T_R} &= \braketi{\bT_R \bT_R}  \propto c_R/2~,\\
\braketi{T_L\bT_L} &\propto (c_L-c_{LR})/2~, \label{TlbTl}\\
\braketi{T_R\bT_R} &\propto (c_R-c_{LR})/2~, \label{TrbTr} \\
\braketi{T_LT_R} &\propto \braketi{\bT_L\bT_R} \propto c_{LR}/2~. \label{TlTr}
\end{align}
\label{TT}
\end{subequations}
For brevity, we omitted the obvious kinematics, which can be inferred from   appendix \ref{app:kinematics}.  
We start with the reflective case: $\ourT_L=\ourT_R=c_{LR}=0.$ Consider the two-point function of the energy flux $T_{xt} \propto T_L-\bT_L = T_R-\bT_R$ on the interface. Due to eqs. (\ref{TlTl},\ref{TlbTl}), it vanishes. Unitarity then implies $T_L=\bT_L$ and $T_R=\bT_R$ as operator equations. This is the converse of the first of the implications in  eq. \eqref{wishlist}:
\beq
\ourT_L=\ourT_R=0 \quad \iff \quad \textup{factorizing interface.}
\eeq
For the maximally transparent case, it is best to distinguish two cases. If $c_L=c_R=c_{LR}$, then $\ourT_L=\ourT_R=1$. This time, we consider the displacement operator \eqref{displacement} $\textup{D}=2( T_L-T_R)$. Due to eqs. (\ref{TlTl},\ref{TlTr}), its two-point function vanishes, and with it the displacement as an operator. We conclude that 
\beq
\ourT_L=\ourT_R=1 \quad \iff \quad c_L=c_R\ \land\ \textup{topological interface.}
\eeq
Finally, if $c_L\neq c_R$, let us take $c_L>c_R$ to fix ideas. Then the maximally transparent interface has $\ourT_L=c_R/c_L$ and $\ourT_R=1$. As first noticed in \cite{Quella:2006de}, this situation is realized by a class of interfaces between two theories with extended symmetries. A subset of the currents of the CFT$_L$ is preserved through the interface and coincides with the full set of currents in the CFT$_R$. Then the interface is completely transparent to the excitations coming from the right. We review this example in more detail in subsection \ref{subsec:cosets}. 

The coefficient $c_{LR}$ has already appeared in connection with the transparency of an interface in \cite{Quella:2006de}. The authors defined the following reflection and transmission coefficients, which are routinely computed to characterize interfaces in $2d$ CFT:
\begin{align}
\quellaT &= \frac{\braketi{T_L(-y)T_R(y)}+\braketi{\bT_L(-y)\bT_R(y)}}{\braket{(T_L(-y)+\bT_R(y))(\bT_L(-y)+T_R(y))}}~, \notag\\
\quellaR &= \frac{\braketi{T_L(-y)\bT_L(-y)}+\braketi{T_R(y)\bT_R(y)}}{\braket{(T_L(-y)+\bT_R(y))(\bT_L(-y)+T_R(y))}}~,
\qquad y>0~.
\label{TRquella}
\end{align}
However, no precise physical interpretation was available for the coefficients \eqref{TRquella}. Now, using eqs. \eqref{TT}, it is easy to see that the coefficients defined in \cite{Quella:2006de} are an average of the left and right transmission and reflection coefficients:
\begin{equation}
\quellaT = \frac{c_L \ourT_L+c_R\ourT_R}{c_L+c_R}~, \qquad
\quellaR = \frac{c_L \ourR_L+c_R\ourR_R}{c_L+c_R}~.
\end{equation}
Notice that the relation is invertible:
\beq
\ourT_L=\frac{c_L+c_R}{2c_L} \quellaT~, \qquad 
\ourT_R=\frac{c_L+c_R}{2c_R} \quellaT~.
\label{quellaVsOur}
\eeq
Despite eq. \eqref{quellaVsOur}, there is a conceptual advantage in $\ourT_{L/R}$ with respect to $\quellaT$. Indeed, $\quellaT$ depends on both the left and the right central charges. On the contrary, $\ourT_L$, say, can be computed by exclusively measuring correlators of the CFT$_L$, as it is physically reasonable -- see eq. \eqref{TlbTl}. For similar reasons, the reflection and transmission coefficients $\ourT_{L/R}$ were already proposed in \cite{Kimura:2015nka}, as a generalization of \cite{Quella:2006de}, in the context of mutliple junctions between critical systems. In fact, it is simple to generalize our analysis to the case of a junction of $n$ wires. The fraction of energy transmitted from wire $i$ to wire $j$ is given by $ c_{ij} / c_i $ where $c_i$ is the central charge of the $i$-th CFT and $c_{ij}$ is the coefficient of the two-point function $\langle T_i T_j \rangle$.

Positivity of the reflected and transmitted energy, as expressed through \eqref{Tbounds}, implies the following bounds on the coefficients \eqref{TRquella}:
\beq
\frac{|c_L-c_R|}{c_L+c_R}\leq \quellaR \leq 1~.
\label{qTunitarity}
\eeq
This bound was conjectured in \cite{Quella:2006de}. On the other hand, a weaker unitarity bound was proven in \cite{Billo:2016cpy}, where the left hand side of \eqref{qTunitarity} appeared squared. The weaker bound followed from reflection positivity applied to the set of two-point functions of the stress tensors of the left and right CFTs. We see that the ANEC tightens the result. Finally, we would like to point out that, through eq. \eqref{QRWcd}, eq. \eqref{qTunitarity} also provides a bound on the two-point function of the displacement operator:
\beq
2 |c_L-c_R|\leq C_\textup{D} \leq 2 (c_L+c_R)~, \qquad
\braketi{\textup{D}(\ii y)\textup{D}(0)}=\frac{C_\textup{D}}{y^4}~.
\label{CDbound}
\eeq 
In the next subsection, we show that the Euclidean bound can be recovered from unitarity of an exclusive process, thus gaining some insight into its relation with the stronger bound coming from the ANEC.

\subsection{Scattering in a defect CFT}
\label{subsec:scattering}

So far, we focused on the expectation value of a flux operator, which is an inclusive observable.
Here we make a few simple observations on an alternative class of exclusive observables. We would like to measure the overlap of two states. The in-state describes the time evolution of a beam which moves towards the interface in the far past. The out-state, when observed in the far future, looks like either a reflected or a transmitted beam. The simplest instance of this process occurs if the in and out states are created by holomorphic or antiholomorphic operators. Let us focus on the stress tensor insertion. We would like to measure the overlap between the following unit normalized states:
\beq
\begin{aligned}
&\textup{in}          &&
                       4 \ep^2\sqrt{\frac{2 }{c_L}}  T_L(z=-\ii \ep)\keti{0} \\
&                            &&
               4 \ep^2\sqrt{\frac{2 }{c_R}} \bT_R(\bz=+\ii \ep) \keti{0}   \\
&\textup{out}      & &
              4 \ep^2\sqrt{\frac{2 }{c_R}}T_R (z=-\ii \ep)\keti{0} \\
&                           &&
4 \ep^2\sqrt{\frac{2 }{c_L}} \bT_L(\bz=\ii \ep) \keti{0}~.
\end{aligned}
\label{inoutbasis}
\eeq
Notice that the in states are orthogonal with each other, and the same is true for the out states.
The overlap of these states defines an $S$ matrix, which is easily computed by means of eq. \eqref{TT}:  
\beq
S=
\begin{pmatrix}
&\frac{c_{LR}}{\sqrt{c_L c_R}}   &    \frac{c_R-c_{LR}}{c_R}  \\
&\frac{c_L-c_{LR}}{c_L}              &   \frac{c_{LR}}{\sqrt{c_L c_R}}
\end{pmatrix}
\eeq
The eigenvalues of $S^\dagger S$ are
\beq
\lambda_1 = 1~, \qquad 
\lambda_2 = \left(1-\frac{c_L+c_R}{c_L c_R} c_{LR}  \right)^2~.
\label{SSdageigen}
\eeq
The presence of a unit eigenvalue is easy to understand: the eigenvector is proportional to the in-state $(T_L+\bT_R)\keti{0}$, which by the gluing conditions eq. \eqref{Tgluing1} is mapped with probability one to the out-state proportional to $(T_R+\bT_L)\keti{0}$. 
The second eigenvalue, instead, imposes the following unitarity bounds:
\beq
0<c_{LR}<2\frac{c_L c_R}{c_L+c_R}~.
\label{clrScatbound}
\eeq
Notice that this condition is strictly weaker than the one found in the main text, eq. \eqref{clrbounds}. This is not surprising, since eq. \eqref{clrbounds} guarantees the positivity of the transmitted and reflected energy in any in-state, while eq. \eqref{clrScatbound} follows from a smaller number of constraints. On the other hand, the bounds \eqref{clrScatbound} are precisely the same as the Euclidean bounds found in \cite{Billo:2016cpy}. This is easily explained, by reinterpreting the overlap of the states \eqref{inoutbasis} in Euclidean signature. The matrix of scalar products among the full set of in and out vectors can be written as follows:
\beq
G=
\begin{pmatrix}
&\mathbb{1}   &S    \\
&S^\dagger    &\mathbb{1}
\end{pmatrix}
\label{Gmatrix}
\eeq
Using properties of block matrices it is easy to prove that each eigenvalue $\lambda_{S^\dagger S}$ of $S^\dagger S$ is related to two eigenvalues $\lambda_{\pm}$ of $G$ as follows:
\beq
\lambda_{\pm} = 1 \pm \sqrt{\lambda_{S^\dagger S}}~.
\eeq
Therefore, the condition $\lambda_{S^\dagger S}\leq 1$ is the same as the constraint that follows from reflection positivity, \emph{i.e.} $\lambda_-\geq 0$. The analysis generalizes to the case where the CFT$_L$ and the CFT$_R$ possess multiple holomorphic and antiholomorphic operators. The $S$ matrix is appropriately enlarged and its eigenvalues provide theory dependent constraints, which are the same as the ones coming from reflection positivity in Euclidean signature.

The Euclidean bounds \eqref{clrScatbound} are in fact equivalent to reflection positivity in a larger set of states. These are generated by the full Verma modules of the following defect Virasoro primaries:
\begin{align}
W_1 =& c_R T_L-c_L \bT_R+c_R \bT_L-c_L T_R~, \\
W_2 =& c_R T_L-c_L \bT_R-c_R \bT_L+c_L T_R~.
\end{align}
All the operators are meant to be evaluated on the defect. $W_1$ and $W_2$ are primaries with respect to the copy of the Virasoro algebra preserved by the defect \cite{Quella:2006de}, and are orthogonal. By requiring their norm to be positive, one easily recovers eq. \eqref{clrScatbound} \cite{Billo:2016cpy}. On the other hand, positivity of the full Verma modules follows from the usual Kac determinant. Indeed, the central charge of the preserved Virasoro algebra is $c_\textup{tot}=c_L+c_R$, and if both CFT$_L$ and CFT$_R$ are unitary, $c_\textup{tot}\geq1$. Standard arguments then show that the representation whose primary has weight 2 is unitary. 

Reflection positivity of these defect modules, together with the one of the identity, can be translated in the unitarity of a larger scattering matrix. Indeed, consider the alternative linear combinations $W=c_R T_L-c_L \bT_R$ and $\overline{W}= c_R \bT_L-c_L T_R$. They are not orthogonal, but they create an in and an out-state respectively, and their descendants preserve the same property.\footnote{One way to see this is to go to the folded picture, \emph{i.e.} to map the defect CFT to a boundary condition for the product theory CFT$_L\times \overline{\textup{CFT}}_R$. The holomorphic/antiholomorphic stress tensors are $T_\textup{tot}= T_L+\bT_R$, $\bT_\textup{tot}= \bT_L+T_R$. Notice that $T_\textup{tot}$ creates an in state and $\bT_\textup{tot}$ an out state. Descendants of $W$ and $\overline{W}$ are obtained by acting with the sum $L^{\textup{tot}}_{-m}+\bar{L}^{\textup{tot}}_{-m}$ in radial quantization around a point on the defect. But $W$ can be thought of as a holomorphic bulk operator, and therefore only $L^{\textup{tot}}_{-m}$ acts on it non trivially. Similarly for $\overline{W}$.} The relation between the scalar products between states in these modules and the $S$ matrix is again as in eq. \eqref{Gmatrix}. By the same argument as before, the $S$ matrix between these in and out-states is unitary. 

One can create more complicated in and out-states by multiple applications of $W$ and $\overline{W}$, or equivalently by acting with other primary operators appearing in the fusion of $W$ and $\overline{W}$ with themselves. It would be interesting to study these more general unitarity constraints. However, at least in the absence of extended symmetry, if the in-state is created by a single stress tensor, then it is orthogonal to all of these more complicated out-states, which belong to different Verma modules on the defect. The discrepancy between the Euclidean bound \eqref{clrScatbound} and the ANEC bound \eqref{clrbounds} then teaches us that the single stress tensor in-state must have a non vanishing overlap with final states which cannot be created by combinations of chiral operators.

One may wonder if the $S$ matrix can be extended to include the overlap between in and out-states created by general non chiral operators. A difficulty immediately arises. Even in the absence of the interface, the overlap between a state created by a local operator in the past and the state created by the same local operator in the future falls off with the distance in time between the two operators.\footnote{Notice that the states should be normalizable. Hence, even when the operators are light-like separated on the Minkowski slice, their square distance is still of order $\epsilon\, T$, $\epsilon$ being the shift in Euclidean time, and $T$ being the time separation between the two insertions.}
The problem arises because of the spread between left and right carriers, which was discussed in section \ref{sec:propagation}. In the far future, the state created by a local operator consists of two beams, moving in opposite directions. If we are only allowed to convolve the operator with a wave packet with compact support, as it happens in the presence of an interface, we cannot project out the left or the right movers. In this sense, the problem with the $S$ matrix is essentially a manifestation of the absence of well defined asymptotic states in an interacting theory with massless degrees of freedom: in the presence of a mass gap, a wave packet with compact support is sufficient to perform the projection onto a well defined single particle state.

The states created by (anti)holomorphic operators are then the only members of a natural basis of in and out-states. Generically, these only amount to the states created by (products of) insertions of the appropriate in or out components of the stress tensor -- see \eqref{inoutbasis}. This set does not form a basis for the Hilbert space of the theory -- for instance, in the absence of an interface they are orthogonal to any state obtained by applying on the vacuum a local operator which is not a Virasoro descendant of the identity. It would be interesting to understand if the basis of in and out-states could be enlarged to contain the chiral parts of every local operator in the theory, perhaps along the lines of \cite{Delfino:2012mx}.

\subsection{Reflection of a global charge}
\label{subsec:global}

It is straightforward to extend the analysis to other conserved quantities.  Here, we sketch the example of the symmetry associated to a global charge. In principle, one could measure the flux of charge at infinity even if the associated symmetry is broken by the interface, but for simplicity we assume that the symmetry is preserved. This request is equivalent to the gluing condition \eqref{gluingJ}, which we repeat here:
\beq
J_L(z)+\O_L\bar{J}_L(\bz)=J_R(z)+\Omega_R\bar{J}_R(\bz)~,\qquad z+\bz=0~,
\eeq
where $\O_L$ and $\O_R$ are automorphisms of the algebra which preserve the stress tensor. 

In this case, the total charge flowing towards the right and towards the left are measured by
\beq
\mathcal{Q}=-\frac{1}{2\pi\ii} \int dz J(z)~, \qquad 
\bar{\mathcal{Q}}=\frac{1}{2\pi\ii} \int d\bz \bar{J}(\bz)~.
\label{Qop}
\eeq
Indeed, consider for instance a quasiprimary operator with charges $(q,\bar{q})$:
\beq
J(z) O(w,\bar{w}) \sim \frac{q}{z-w} O(w,\bar{w}) +\textup{regular}~,
\eeq
and similarly for $\bar{J}$. It is then immediate to see that the expectation value of the operators \eqref{Qop} in a state created by $O$ is the correct one:\footnote{Of course, the bra $\bra{O}$ is obtained by inserting the operator $O^\dagger$, which has charge $-q$.}
\beq
\frac{\braket{O|\mathcal{Q}|O}}{\braket{O|O}}=q~,
\qquad
\frac{\braket{O|\bar{\mathcal{Q}}|O}}{\braket{O|O}}=\bar{q}~.
\eeq
The definition of the reflection and transmission coefficients for the global charge is analogous to eq. \eqref{TRenergyhhb}, and the rest of the discussion goes through unchanged as well. The transmission coefficient in a state created by a charged local operator on the left of the interface is determined by the coupling of the $(1,0)$ quasi-primaries of the CFT$_L$ with $J_R$. If $J_L$ is the only current which couples to $J_R$, then the transmission coefficients are
\beq
\mathcal{T}^{\mathcal{Q}}_L=\frac{k_{LR}}{k_L}~, \qquad \mathcal{T}^{\mathcal{Q}}_R=\frac{k_{LR}}{k_R}~.
\label{QTR}
\eeq
Here we defined the following quantities:
\beq
\braket{J_L(z) J_L(z')}=\frac{k_L}{(z-z')^2}~, \qquad \braket{J_L(z) J_R(z')}=\frac{k_{LR}}{(z-z')^2}~.
\eeq
We further assumed that $k_L=\bar{k}_L$, and similarly for $k_R$. The charge transmission coefficient defined in \cite{Kimura:2014hva}, which was derived by analogy with \cite{Quella:2006de}, is, again, an average of the coefficients \eqref{QTR}. Notice, however, that when $J_L$ -- say -- is part of a non-abelian symmetry algebra, the CFT$_L$ contains in general other currents which commute with $J_L$. Depending on the specific boundary conditions, they could couple with $J_R$. The transmission and reflection coefficients may then depend in a more detailed way on the weight vector describing the representation of the operator which creates the state. It would be interesting to analyze this scenario in detail.

\section{Examples}
\label{sec:examples}

One of the nice features of the result \eqref{TLsl2general} is that in many cases reflection and transmission are exactly computable. Although we shall not need it in the following, let us briefly review a general procedure to compute the crucial coefficient $c_{LR}$, following \cite{Quella:2006de}. To each conformal interface between the theories CFT$_L$ and CFT$_R$, one associates a conformal boundary condition for the theory CFT$_L \times \overline{\textup{CFT}}_R$ obtained by applying a parity transformation $x\to -x$ to the CFT$_R$. This is the so-called folding trick. The parity transformation interchanges holomorphic and antiholomorphic operators, so in particular $\tilde{T}_R(\bz)\equiv T_R(z')$, with $z'=-\bz$, is a $(0,2)$ operator of the $\overline{\textup{CFT}}_R$. The knowledge of the Cardy state $\ket{b}$ associated to this boundary condition is equivalent to the knowledge of all the one-point functions of local operators in the presence of the interface.  The coefficient $c_{LR}$ is then computed as the one-point function of the quasi-primary $T_L\tilde{T}_R$:
\beq
c_{LR}= 2 \braket{T_L\tilde{T}_R|b}~.
\eeq
The authors of \cite{Quella:2006de} used this method to compute the averaged coefficients $\quellaT$ and $\quellaR$ defined in \eqref{TRquella} for a variety of defect CFTs with the feature that the folded theory CFT$_L \times \overline{\textup{CFT}}_R$ is rational. Of course, the reflection and transmission coefficients defined in this paper can be easily extracted from their results. The same is true for the interfaces considered in various later works -- see e.g. \cite{Gang:2008sz,Kimura:2015nka,Brunner:2015vva,Makabe:2017ygy}. We refer to \cite{Quella:2006de} for further details, and in the following we focus instead on some very simple examples in order to check and exemplify  the universality of the result \eqref{TLsl2general}.

\subsection{The free boson}
\label{subsec:boson}

The theory of a single free boson $\phi$ admits a simple family of conformal interfaces, which was constructed in \cite{Bachas:2001vj}. The interface is completely specified by a gluing matrix:
\beq
\begin{pmatrix}
\pa \phi_L \\
\bar{\pa} \phi_R 
\end{pmatrix}
= S_\pm
\begin{pmatrix}
\bar{\pa} \phi_L \\
\pa \phi_R 
\end{pmatrix}~,
\qquad 
S_\pm=
\begin{pmatrix}
\cos 2\theta & \pm\sin 2\theta \\
\sin 2\theta  & \mp \cos 2\theta
\end{pmatrix}~.
\label{gluingS}
\eeq
Here, all operators are evaluated at the interface $z=-\bz$, although the equation can be analytically continued inside correlation functions. The matrix $S$ is labeled by an angle $\theta$, and is fixed by requiring the gluing condition \eqref{gluingT} to hold. Notice that the two choices in eq. \eqref{gluingS} are obtained one from the other by acting with T-duality on one side only, \emph{e.g.} $\pa\phi_R\to - \pa\phi_R$, $\bar{\pa}\phi_R\to \bar{\pa}\phi_R$. Below, we choose the upper sign, \emph{i.e.} $S_+$. It is convenient to think of the boson $\phi_L$ as compactified with a radius $R_L$, while the radius of compactification of $\phi_R$ is $R_R$. Then $\tan \theta =   q \frac{ R_L}{ R_R}$ for some rational $q$, so that the gluing condition preserves the periodicities. 

From eq. \eqref{gluingS} we get
\beq
\ourT_L=\ourT_R =c_{LR}= (\sin 2\theta)^2~.
\label{Tfree}
\eeq
The universality of the coefficient is physically clear: the $S$ matrix fixes the amplitude for a particle to be transmitted, which means that the transmitted energy in any multiparticle state is fixed as well. Technically, the same result easily follows from Wick theorem. From the gluing condition eq. \eqref{gluingS}, one gets the following Euclidean propagators:
\begin{align}
\braketi{\phi_L(z_1,\bz_1)\phi_L(z_2,\bz_2)} &= -\log |z_1-z_2|^2
+\cos 2\th \log |z_1+\bz_2|^2~, \notag \\
\braketi{\phi_R(z_1,\bz_1)\phi_R(z_2,\bz_2)} &= -\log |z_1-z_2|^2
-\cos 2\th \log |z_1+\bz_2|^2~, \label{freeprop}\\
\braketi{\phi_L(z_1,\bz_1)\phi_R(z_2,\bz_2)} &= -\sin 2\th \log|z_1-z_2|^2~. \notag
\end{align}
The most general local operator on the left of the interface is a linear combination of products of the following building blocks: a vertex operator $V_{k,\bar{k}}(z,\bz)$ with dimensions $(k^2/2,\bar{k}^2/2)$, and derivatives of the field $\phi_L$. The bulk operators are defined in the translational invariant theory, therefore they are normal ordered by only subtracting the singular part of the propagators \eqref{freeprop}. It is convenient to also define a different normal ordering -- which we denote $:\ :$ -- by subtracting the complete propagator. When the operator $:V_{k,\bar{k}}(z,\bz):$ is inserted in a correlator, the only contractions left to be performed are with other operators in the correlator. In particular, it is easy to verify that
\beq
V_{k,\bar{k}}(z_1,\bz_1) V_{-k,-\bar{k}}(z_2,\bz_2) = 
\frac{(\xi+1)^{\cos2\th\, k \bar{k}}}{(z_1-z_2)^{k^2}(\bz_1-\bz_2)^{\bar{k}^2}}
: V_{k,\bar{k}}(z_1,\bz_1) V_{-k,-\bar{k}}(z_2,\bz_2)  :
\label{normord}
\eeq
Since the interface preserves the target $U(1)$ symmetry, the most general three-point function which we need to consider is the following:
\beq
\braketi{V_{k,\bar{k}}(z_1,\bz_1)O_1(z_1,\bz_1)\, T_R(z)\, V_{-k,-\bar{k}}(z_2,\bz_2)O_2(z_2,\bz_2)}~,
\eeq
where $O_1$ and $O_2$ are products of derivatives of $\phi_L$ and $T_R=1/2\,\pa \phi_R\pa \phi_R$. After plugging in eq. \eqref{normord}, all the remaining contractions involve derivatives of the propagators \eqref{freeprop}. As we send $D\to \infty$, $\xi \to 0$ and the prefactor in eq. \eqref{normord} simplifies. Moreover, all the additional contractions involving $(z_1+\bz_2)$ and $(z_2+\bz_1)$ are suppressed. The result equals the correlator obtained by removing the interface and replacing $T_R \to T_L$, except for a factor $(\sin 2\theta)^2$, which arises from the prefactor in the third line of eq. \eqref{freeprop}. This explicitly verifies that eq. \eqref{Tfree} gives the transmission coefficient for any state in the theory.

\subsection{The free fermion and the Ising model}
\label{subsec:ising}

The theory of a single Majorana fermion admits a class of defects analogous to the one discussed for the free boson: an amplitude for transmission and reflection is assigned to left and right movers. In this case, the gluing condition \eqref{gluingT} imposes the following form for the $S$ matrix:
\beq
\begin{pmatrix}
\psi_L \\
\bar{\psi}_R 
\end{pmatrix}
= S_\pm
\begin{pmatrix}
\bar{\psi}_L \\
\psi_R 
\end{pmatrix}~,
\qquad 
S_\pm=
\begin{pmatrix}
\ii \sin \chi & \pm\cos \chi \\
\cos \chi  & \pm \ii\sin \chi
\end{pmatrix}~.
\label{gluingSfer}
\eeq
The angle $\chi$ is fixed to be real by the reality conditions $\psi^\dagger=\ii \psi$, $\bar{\psi}^\dagger=-\ii \bar{\psi}$. Also in this case, the two families of interfaces are related by the chiral transformation $\psi_R \to - \psi_R$, $\bar{\psi}_R \to  \bar{\psi}_R$, which in the Ising description corresponds to applying Kramers-Wannier duality to the CFT$_R$ only. In the following, we choose $S_+$ in eq. \eqref{gluingSfer}, that is, we choose the family which is connected with the trivial defect at $\chi=0$. This family of defects was considered in \cite{1994PhLB..328..123D}.

Multiparticle states can be proven to have the same reflection coefficient via Wick theorem, analogously to what has been done in the previous subsection. A trivial computation yields
\beq
\ourT_L=\ourT_R =2 c_{LR}= (\cos \chi)^2~.
\label{Tfreefer}
\eeq

The Ising model has a description in terms of a free Majorana fermion. However, the primary $\sigma$, with dimensions $(1/16,1/16)$, is not a local operator of the free fermion theory, and by applying it on the vacuum we gain access to a state which cannot be expressed as a superposition of a finite number of particles. It is interesting to check that eq. \eqref{Tfreefer} provides the correct transmission coefficient also in this case. The conformal interfaces which can be embedded in the Ising model are known \cite{Oshikawa:1996dj} thanks to the fact that the folded theory is equivalent to an orbifolded free boson.\footnote{The known interfaces in the Ising model correspond to Neumann and Dirichlet boundary conditions for the free boson. There is however a third class of boundary conditions for the free boson \cite{Friedanc1,Janik:2001hb}, whose corresponding defect in the Ising model has not been studied.} The reflection and transmission coefficients for these interfaces are given in \cite{Quella:2006de}. The authors also provide the map between the $S_+$ family in eq. \eqref{gluingSfer} and a family of defects in the Ising model. 

\begin{figure}[]
\centering
\begin{tikzpicture}[scale=1.2]
\draw [decorate,decoration={snake,amplitude=0.8pt}] (-1,1.5) .. controls (-2,1) and (-2.5,-0.5) .. (-1.5,-1);
\draw [decorate,decoration={snake,amplitude=0.8pt}] (1,1.5) .. controls (2.3,1) and (2.8,-0.5) .. (1.5,-1);
\filldraw [black] (-1,1.5) circle [radius=1pt] node[above] {$z_1$};
\filldraw [black] (-1.5,-1) circle [radius=1pt] node[above] {$z_2$};
\filldraw [black] (1,1.5) circle [radius=1pt] node[above] {$-\bz_1$};
\filldraw [black] (1.5,-1) circle [radius=1pt] node[above] {$-\bz_2$};
\filldraw [black] (1.3,0) circle [radius=1pt] node[above] {$z$};
\draw [postaction = {decoration={markings, mark=between positions 0.2 and 0.9 step 1cm  with {\arrow{>};}},decorate}] (0,-2) -- (0,2);
\draw [black] (3,1.8)-- (2.8,1.8) -- (2.8,2);
\node at (2.95,1.91) {$w$};
\end{tikzpicture}
\caption{The contour of integration in eq. \eqref{IsingG1} runs on the imaginary axis and is marked by arrows. The position of the singularities of the integrand refers to a Euclidean configuration.}
\label{fig:isingg}
\end{figure}
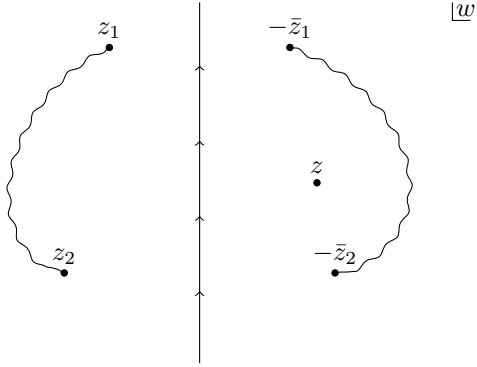

Although it would be interesting to compute the exact three point function $\braketi{\si_L T_R \si_L}$, we shall content ourselves with a perturbative computation around the trivial interface at $\chi=0$. In this setup, it is easy to see how to construct the $S_+$ interface \cite{PhysRevLett.44.840,PhysRevB.25.331}. The energy field $\ep=\ii \psi\bar{\psi}$ has weight $( \frac{1}{2},\frac{1}{2} )$, and can be used to generate a one parameter family of interfaces, by inserting in the path integral the following operator: 
\beq
\mathcal{D}= \exp \left\{-\frac{g}{2\pi} \int_{-\infty}^{+\infty}\! d\tau\, \ep(x=0,\tau)\right\}~.
\label{IsingD}
\eeq
We match $g$ to $\chi$ by computing the two-point functions of the fermions at the leading non trivial order:
\beq
\chi = g+O(g^3)~.
\eeq
We are now ready to compute the transmission coefficient in the state created by $\si$, at lowest order in $g$. 
We start in a Euclidean configuration, with the interface on the imaginary axis. We expect the first non trivial contribution to $\ourT_L$ to arise at order $g^2$, therefore we define
\begin{multline}
\braketi{\si_L(z_1,\bz_1)T_R(z)\si_L(z_2,\bz_2)} \\
= G_0(z_1,z_2,\bz_1,\bz_2,z) +
\frac{g}{2\pi}\, G_1(z_1,z_2,\bz_1,\bz_2,z)+ \left(\frac{g}{2\pi}\right)^2 \, G_2(z_1,z_2,\bz_1,\bz_2,z)+\dots
\end{multline}
The zeroth order is simply given by the correlator without interface,
\beq
G_0(z_1,z_2,\bz_1,\bz_2,z)=\braket{\si(z_1,\bz_1)T(z)\si(z_2,\bz_2)}~.
\eeq
Although we do not expect a contribution to $\ourT_L$ at order $g$, it is interesting to consider the leading order as an example of the structure of singularities in a defect CFT:\footnote{Notice that from here on the arguments of the operator $\ep$ are the usual holomorphic and antiholomorphic coordinates, contrary to eq. \eqref{IsingD}, where we used Cartesian coordinates.}
\beq
G_1(z_1,z_2,\bz_1,\bz_2,z) =  \ii \int_{-\ii\infty}^{+\ii\infty}\!\!dw 
\braket{\si(z_1,\bz_1)T(z)\si(z_2,\bz_2)\ep( w,-w)}~.
\label{IsingG1}
\eeq
The integrand is\footnote{It is possible to recast the integral \eqref{IsingG1} in the following covariant form:
\begin{multline}
G_1(z_1,z_2,\bz_1,\bz_2)=\frac{1}{(z-z_1)^2(z-z_2)^2(z_1-z_2)^{-15/8}(\bz_1-\bz_2)^{1/8}} \\
\times\ii\int_\mathcal{C}dw\left(\frac{1}{16\sqrt{w}}+\frac{(1-u)\sqrt{w}}{2 \left(1+(u-1)w\right)^2}\right) \sqrt{\frac{\xi}{(w-1)\left((1+\xi)w-1\right)}}~,
\end{multline}
where the contour $\mathcal{C}$ runs upwards, and leaves to the left the cut joining $w=0$ to $w=\infty$, and to the right both the cut which joins $w=1$ to $w=1/(1+\xi)$ and the pole at $w=1/(1-u)$.
}
\begin{multline}
\braket{\si(z_1,\bz_1)T(z)\si(z_2,\bz_2)\ep(w,- w)} = 
\left(\frac{1/16}{(z-z_1)^2}+\frac{1/16}{(z-z_2)^2}+\frac{1}{z-z_1}\frac{\pa}{\pa z_1}+\frac{1}{z-z_2}\frac{\pa}{\pa z_2}\right)\braket{\si_1\si_2\ep} \\
+\left(\frac{1/2}{(z- w)^2}-\frac{1}{z- w} \left(\frac{\pa}{\pa z_1}+\frac{\pa}{\pa z_2}\right) \right) \braket{\si_1\si_2\ep}~, \label{IsinggIntegrand}
\end{multline}
where we used the following shorthand notation:
\beq
\braket{\si_1\si_2\ep} 
=c_{\sigma\sigma\epsilon} 
\frac{[(z_1-z_2)(\bz_1-\bz_2)]^{3/8}}{[(z_1-w)(\bz_1+w)(z_2-w)(\bz_2+w)]^{1/2}}~.
\label{threepointising}
\eeq
We will not need the value of $c_{\sigma\sigma\epsilon}$.
The contour and the singularities in the $w$ plane are shown in fig. \ref{fig:isingg}.
 It is important that $\Re z_1,\Re z_2<0$, and $\Re z>0$. It is possible to analytically continue the correlator as a function of $z_i,\bz_i$ and $z$, before integrating in $w$, as long as we are careful to deform the contour of integration such that $z_1,\,z_2$ lie on the left, and $-\bz_1,\,-\bz_2,\,z$ lie on the right. Hence, we can study the nature of the singularities as $z$ collides with other insertions without performing the integral. There are no singularities as $z\to \bz_i$, as expected. On the contrary, there are branch cuts originating from $z=z_1$ and $z=z_2$, which correspond to $u=1$ and $u=\infty$ in fig. \ref{fig:crosscontour}. Indeed, when we move $z$ to the region on the left of the contour, we pick up the contribution of the pole at $w=z$ in the second line of eq. \eqref{IsinggIntegrand}:
\begin{multline}
\left. G_1(z_1,z_2,\bz_1,\bz_2,z) \right|_{\Re z<0} = \ii \int_{-\ii\infty}^{+\ii\infty}\!\!dw 
\braket{\si(z_1,\bz_1)T(z)\si(z_2,\bz_2)\ep( w,-w)} \\
+2\pi\, \textup{Res}_{w=z} \left\{ \left(\frac{1/2}{(z- w)^2}-\frac{1}{z- w} \left(\frac{\pa}{\pa z_1}+\frac{\pa}{\pa z_2}\right) \right) \braket{\si_1\si_2\ep} \right\}~.
\end{multline}
While the first line in the previous equation has no monodromies as $z$ goes around $z_1$ or $z_2$, the second line has square root branch points, as it is immediate to verify using eq. \eqref{threepointising}.

We shall not attempt to compute $G_1$ explicitly. Rather, in appendix \ref{app:ising} we show that the result does not contribute to the transmission coefficient.

\begin{figure}[]
\centering
\begin{tikzpicture}[scale=1.2]
\draw [decorate,decoration={snake,amplitude=0.8pt}] (-1.5,1.5) .. controls (-2.5,1) and (-3,-0.5) .. (-2,-1);
\draw [decorate,decoration={snake,amplitude=0.8pt}] (1.7,1.5) .. controls (2.7,1) and (3.5,-0.5) .. (2.2,-1);
\draw [postaction = {decoration={markings, mark=between positions 0.2 and 0.9 step 0.3  with {\arrow{<};}},decorate}]  (-0.6,0) circle [radius=0.32];
\draw [postaction = {decoration={markings, mark=between positions 0.2 and 0.9 step 0.3  with {\arrow{<};}},decorate}]  (-0.6,0) circle [radius=0.62];
\filldraw [black] (-1.5,1.5) circle [radius=1pt] node[above] {$z_1$};
\filldraw [black] (-2,-1) circle [radius=1pt] node[above] {$z_2$};
\filldraw [black] (1.7,1.5) circle [radius=1pt] node[above] {$-\bz_1$};
\filldraw [black] (2.2,-1) circle [radius=1pt] node[above] {$-\bz_2$};
\filldraw [black] (-0.6,0) circle [radius=1pt] node[above] {$z$};
\draw[->] (2,1.5) -- (2.8,1.5);
\draw[->] (2.4,-1) -- (3.2,-1);
\draw [black] (4,1.8)-- (3.8,1.8) -- (3.8,2);
\node at (3.92,1.91) {$w$};
\end{tikzpicture}
\caption{The deformed contours of integrations in $w_1$ and $w_2$ which give a nonvanishing contribution to $G_2$ in the $D\to\infty$ limit. The effect of the limit is to carry the points $-\bz_1$ and $-\bz_2$ to infinity, as shown by the arrows.}
\label{fig:isingresidues}
\end{figure}
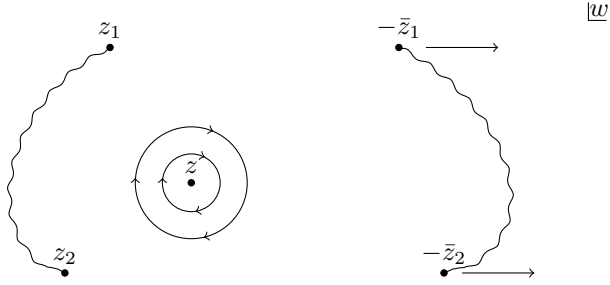

Let us move on to the order $g^2$:
\begin{multline}
G_2(z_1,z_2,\bz_1,\bz_2,z)=-\frac{1}{2} \int_{-\ii\infty}^{+\ii\infty}\!\!dw_1 dw_2 
\braket{\si(z_1,\bz_1)T(z)\si(z_2,\bz_2)\ep( w_1,-w_1)\ep( w_2,-w_2)} \\
+\frac{1}{2} \braket{\si(z_1,\bz_1)T(z)\si(z_2,\bz_2)} \int_{-\ii\infty}^{+\ii\infty}\!\!dw_1 dw_2 
\braket{\ep( w_1,-w_1)\ep( w_2,-w_2)}~.
\label{IsingG2}
\end{multline}
The relevant five-point function can be easily found as in eq. \eqref{IsinggIntegrand}, in terms of the following four-point function \cite{Belavin:1984vu}:
\beq
\braket{\si(z_1,\bz_1) \si(z_2,\bz_2)\ep( w_1,\bar{w}_1)\ep( w_2,\bar{w}_2)} = 
\frac{F(x)F(\bar{x})}{(z_{12}\bz_{12})^{1/8}w_{12}\bar{w}_{12}}~, \quad F(x)=\frac{2-x}{2\sqrt{1-x}}~,
\eeq
where
\beq
x=\frac{(z_1-z_2)(w_1-w_2)}{(z_1-w_1)(z_2-w_2)}~.
\eeq
Let us first notice that a power law singularity arises in eq. \eqref{IsingG2} as $w_1\to w_2$, due to the fusion rule $\ep\times \ep \sim 1$. This divergence is harmless and can be ignored. We refer to the appendix for an explicit demonstration of this fact. We would like to extract the $D\to\infty$ limit of the integral \eqref{IsingG2}. Let us focus on the connected contribution. The position of the singularities in the $w_1$ and $w_2$ planes is the same as in fig. \eqref{fig:isingg}. Our strategy is the following: we shift both contours of integration to the right of the $z$ insertion, and we keep them at a distance of order $D \simeq |z_1+\bz_1|$ from both $z_i$ and $\bz_i$ as $D$ grows. It is clear that, unless we pick the residues at the pole $w_i=z$, the two insertions of $\epsilon$ factorize in the limit, and this contribution cancels against the second line in eq. \eqref{IsingG2}. It is also easy to see that the term where we only pick the pole in $z$, say, in the $w_1$ integral, falls off as $D \to \infty$. The only non vanishing contribution then comes from integrating along the contours shown in fig. \ref{fig:isingresidues}.
The result is 
\begin{multline}
\lim_{D\to\infty}\braketi{\si_L(z_1,\bz_1-D)T_R(z)\si_L(z_2,\bz_2-D)}
= (1-g^2+ \dots) \frac{1/16}{\bz_{12}^{1/8}z_{12}^{-15/8}(z-z_1)^2(z-z_2)^2}\\
= (1-g^2+ \dots) \braket{\si(z_1,\bz_1)T(z)\si(z_2,\bz_2)}~.
\label{isingDinfty}
\end{multline}
By comparing with \eqref{Tfreefer}, we see that eq. \eqref{isingDinfty} reproduces the general result \eqref{OOTxizero}. It also establishes that, as expected, the transmission coefficient in states created by the spin field or by a finite number of free fermions are equal.

As a final check, in appendix \ref{app:ising} we show that the $D\to \infty$ limit commutes with the integration of the stress tensor, by explicitly performing the integration first.

\subsection{An example with multiple operators of weight $(2,0)$}
\label{subsec:cosets}

If a CFT is endowed with an extended symmetry, the holomorphic stress tensor is generically not the only $(2,0)$ quasi-primary. Hence, permeable interfaces between theories of this kind provide examples where the reflection and transmission coefficients are not fully determined by $c_{LR}$, and in particular depend on the local operator used to create the state. Familiar instances of theories with extended symmetry are the rational CFTs whose chiral algebra $\mathcal{A}(\mathfrak{g}_k)$ is generated by the currents of an affine Lie Algebra $\mathfrak{g}$ at level $k$.
 In \cite{Quella:2002ct}, a procedure was devised to construct rational interfaces between theories whose partition function is the charge conjugated modular invariant of two chiral algebras $\mathcal{A}_L$ and $\mathcal{A}_R$. 
Here, we shall consider a simple example, which is sufficient to illustrate the new features of this setup.

Consider a WZW model based on the $su(2)_k$ chiral algebra. There is an interface between this theory -- the CFT$_L$, for definiteness -- and a compact free boson with radius $R=\sqrt{2k}$, which we denote as $u(1)_k$ -- the CFT$_R$. Indeed, $u(1)_k$ is an affine subalgebra of $su(2)_k$, and the integrable representations of $su(2)_k$ can be decomposed in finitely many representations of the (extended) algebra generated by the currents of  $u(1)_k$ \cite{DiFrancesco:1997nk}. The technology of \cite{Quella:2002ct} allows to construct a rational interface which preserves the reduced chiral algebra $su(2)_k/u(1)_k\oplus u(1)_k$ out of the full $su(2)_k$ of the CFT$_L$. The currents which generate the extended symmetry of the $u(1)_k$ subalgebra are glued with totally transmittive boundary conditions, $J_L=J_R$, while the currents in the coset $su(2)_k/u(1)_k$ -- for generic $k$ only the stress tensor --  obey the factorizing boundary condition \eqref{factorizing}. 

We should expect the energy reflection and transmission coefficients to be able to distinguish between the two kinds of degrees of freedom.\footnote{In fact, the coset $su(2)_k/u(1)_k$ has an interpretation in terms of parafermions \cite{DiFrancesco:1997nk}. The interface is then factorizing for the parafermions, and topological for the bosons.} Let us see that this is indeed the case. $\ourT_L$ is determined by the set of $(2,0)$ left quasi-primaries which couple with the stress tensor of the CFT$_R$. From the $su(2)_k$ characters, we see that for a generic value of $k$ the CFT$_L$ contains two $(2,0)$ quasi-primaries which are neutral under the $u(1)$ generated by $j^3$, the generator of the Cartan subalgebra. We can identify these operators with the stress tensors of the coset $T^{su(2)/u(1)}$ and of the $u(1)$ subalgebra $T^{u(1)}$. Since their modes commute, they are orthogonal in the homogeneous theory. Their sum is the stress tensor of the CFT$_L$. The gluing conditions precisely state that $T^{su(2)/u(1)}$ does not correlate with $T_R$, while $T^{u(1)}$ is identified with it:
\beq
T^{su(2)/u(1)}=\bT^{su(2)/u(1)}~,
\qquad
T^{u(1)} = T_R~,
\qquad\quad \textup{at the interface,}
\eeq
which means
\beq
\braketi{T^{su(2)/u(1)}(z)T_R(z')}=0~,
\qquad
\braketi{T^{u(1)}(z) T_R(z')} = \frac{c_{u(1)}/2}{(z-z')^4}~.
\eeq
Here, $c_{u(1)}=1$ is the central charge of the free boson. As an immediate consequence, $\ourT_L=1$ in all the states created by operators which project to the identity field in the coset. Indeed, their self OPE only contains $T^{u(1)}$ at level $(2,0)$.  These include of course all the states created by the modes of the $u(1)_k$ currents out of the vacuum.\footnote{The identity representation of the coset theory appears in two representations of $su(2)_k$: the identity and, due to field identifications \cite{DiFrancesco:1997nk}, the representation whose primary has maximal $su(2)$ spin, \emph{i.e.} $k/2$.} At the opposite extreme, the transmission coefficient vanishes in the states created by the modes of the coset stress tensor $T^{su(2)/u(1)}$. More generally, let us consider an operator $O_L$ of $su(2)_k$ which is both a quasi-primary under $T^{su(2)/u(1)}$ and under $T^{u(1)}$. The simplest example is the current $j^+$. Then the relevant part of the fusion rules read
\beq
O_L \times O^\dagger_L \sim \frac{2h_{su(2)/u(1)}}{c_{su(2)/u(1)}}T^{su(2)/u(1)}+\frac{2h_{u(1)}}{c_{u(1)}}T^{u(1)}~,
\eeq
where the meaning of the subscript in the holomorphic weights and the central charges is the obvious one. It follows that the transmission coefficient in the state created by $O_L$ is
\beq
\ourT(O_L)=\frac{h_{u(1)}}{h_{su(2)}}~,
\eeq
where the denominator is the holomorphic weight as measured by the stress tensor $T_L$ of the complete $su(2)_k$ theory. For instance, $\ourT(j^+)=1/k$. In particular, the transmission coefficients become unity when $k=1$, since in that case the free boson theory on the right, which is at the self-dual radius, is precisely the same as the $su(2)_1$ WZW model on the left.

As mentioned in subsection \ref{subsec:discussion}, $c_{LR}$ still determines the reflection coefficient in the states created by the modes of the stress tensor $T_L$. Indeed, the only $(2,0)$ quasi-primary appearing in the OPE of $T_L$ with itself is again the stress tensor.  
Hence, eq. \eqref{TLsl2general} is always valid at least in this subset of the states of the defect CFT, and with it the bounds discussed in subsection \ref{subsec:bounds}. Here,
\beq
c_{LR} = c_{u(1)}~, \qquad \ourT(T_L)=\frac{1}{c_{su(2)}}=\frac{k+2}{3k}~.
\eeq

Finally, it is easy to check that the energy coming from the right is completely transmitted, \emph{i.e.}
\beq
\ourT_R=1.
\eeq

We refrain from analysing more general interfaces built with the method of \cite{Quella:2002ct}, but let us at least emphasize some features which are common to all the cases where the CFT$_L$ and the CFT$_R$ are generated by two affine algebras $\mathcal{A}_L$ and $\mathcal{A}_R$, with a common subalgebra $\mathcal{C}$ to which the interface is transparent. First of all, the transmission coefficient $\ourT=1$ for the states created by the generators of $\mathcal{C}$. On the contrary, $\ourT=0$ for the generators of the cosets $\mathcal{A}_L/\mathcal{C}$ and $\mathcal{A}_R/\mathcal{C}$, \emph{i.e.} the currents which commute with the generators of $\mathcal{C}$.\footnote{The proof of that $\ourT=1$ for elements of $\mathcal{C}$ goes as follows. Given the currents $J^a \in \mathcal{C}$, the commutation relations imply that the non singular term in the OPE $J^a \times J^b$ corresponds, up to a descendant, to the state $J_{-1}^aJ_{-1}^b\ket{0}$. When rearranging in irreducible representations of $\mathcal{C}$, the only neutral operator is the Casimir, \emph{i.e.} the Sugawara $T^{\mathcal{C}}$. The result then follows by comparing, for instance, $\braketi{T_L^{\mathcal{C}}T_R}$ with $\braket{T_L^{\mathcal{C}}T_L}$. Instead, the proof that $\ourT=0$ for the generators $J_L$ of the coset $\mathcal{A}_L/\mathcal{C}$ follows from the fact that these generators obey the factorizing boundary conditions $J_L+\bar{J}_L=0$. It is then easy to see that the three-point function $\braketi{J_LT_RJ_L}$ vanishes.} Finally, as first noticed in \cite{Quella:2006de}, the coefficient $c_{LR}$ always coincides with the central charge of $\mathcal{C}$.

\section{Non-equilibrium steady states}
\label{sec:steady}

The general result \eqref{mainresult} has a simple and powerful consequence on the physics of non-equilibrium steady states at criticality. These are systems characterized by non vanishing fluxes which are however time independent, see  \cite{Bernard:2016nci} for a review.  Consider a wire which is well described by a 1+1 dimensional CFT.
In thermal equilibrium, there is the same energy flux of left and right movers along the wire, given by
\beq
\langle T(z) \rangle =\langle \bar{T}(\bar{z}) \rangle = -\frac{\pi^2 c}{6 \beta^2}
\eeq
where $c$ is the central charge and $\beta$ is the inverse temperature. 
Thus, the total energy density and energy flux are given by 
\beq
\langle T^{tt}\rangle=-\frac{1}{2\pi}\left(\langle T(z)\rangle+\langle \bar{T}(\bar{z})\rangle \right)=  \frac{\pi c}{6 \beta^2}\,,\qquad \qquad
\langle T^{tx}\rangle=-\frac{1}{2\pi}\left(\langle T(z)\rangle- \langle\bar{T}(\bar{z})\rangle \right)= 0\,.
\eeq

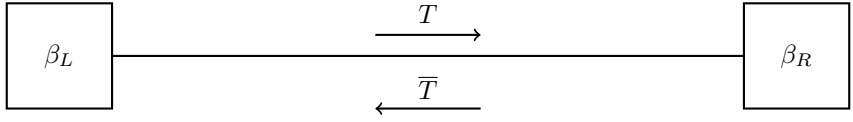
\begin{figure}
\centering
\begin{tikzpicture}[scale=1.4]
\draw [thick] (-4,-0.5) rectangle (-3,0.5) node at (-3.5,0) {$\beta_L$};
\draw [thick] (3,-0.5) rectangle (4,0.5) node at (3.5,0) {$\beta_R$};
\draw [thick] (-3,0) -- (3,0);
\draw [thick,->] (-0.5,0.2) -- (0.5,0.2) node at (0,0.4) {$T$};
\draw [thick,->] (0.5,-0.5) --  (-0.5,-0.5) node at (0,-0.3) {$\bT$};
\end{tikzpicture}
\caption{Depiction of a quantum wire in a non-equilibrium steady state. The thermal reservoirs are denoted as boxes.}
\label{fig:wire1}
\end{figure}
On the contrary, if we place two ends of the wire at different temperatures as in figure \ref{fig:wire1}, energy flows from one reservoir to the other. Since in CFT left-moving and right-moving excitations do not see each other, the energy flux in steady state is given by 
\beq
\langle T^{tx}\rangle=-\frac{1}{2\pi}\left(\langle T(z)\rangle- \langle\bar{T}(\bar{z})\rangle \right)=  \frac{\pi c }{12  } \left( \frac{1}{  \beta_L^2} - \frac{  1}{ \beta_R^2} \right)\,.
\label{currentnodefect}
\eeq
Notably, the energy flux between the thermal reservoirs does not depend on the length of the wire connecting them. Such ballistic transport has been studied in \cite{Bernard:2012je,Bernard:2014fca}.

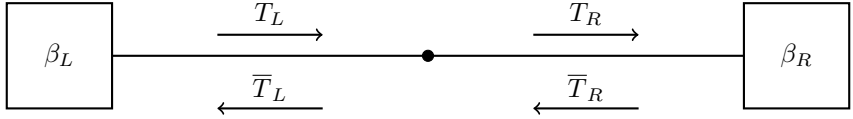
\begin{figure}
\centering
\begin{tikzpicture}[scale=1.4]
\draw [thick] (-4,-0.5) rectangle (-3,0.5) node at (-3.5,0) {$\beta_L$};
\draw [thick] (3,-0.5) rectangle (4,0.5) node at (3.5,0) {$\beta_R$};
\draw [thick] (-3,0) -- (3,0);
\filldraw [black] (0,0) circle [radius=1.5pt];
\draw [thick,->] (-2,0.2) -- (-1,0.2) node at (-1.5,0.4) {$T_L$};
\draw [thick,->] (-1,-0.5) --  (-2,-0.5) node at (-1.5,-0.3) {$\bT_L$};
\draw [thick,->] (1,0.2) -- (2,0.2) node at (1.5,0.4) {$T_R$};
\draw [thick,->] (2,-0.5) --  (1,-0.5) node at (1.5,-0.3) {$\bT_R$};
\end{tikzpicture}
\caption{Same as in fig. \ref{fig:wire1}, with an impurity depicted as a black circle.}
\label{fig:wire2}
\end{figure}

Let us now consider two wires connected by a conformal interface and attached to thermal reservoirs as in figure \ref{fig:wire2}. If there is a unique $(2,0)$ current, then
given the universality of the transmission coefficient and the decoherent  nature of thermal radiation we conclude that
\begin{align}
\langle T_L(z)\rangle &= - \frac{\pi c_L}{6 \beta_L^2} \,, \qquad
\qquad 
\langle 
\bar{T}_L(\bar{z})\rangle = \langle \bar{T}_R(\bar{z})\rangle \mathcal{T}_R +\langle T_L(z)\rangle \mathcal{R}_L \,,\\
\langle \bar{T}_R(\bar{z})\rangle &=-  \frac{\pi c_R}{6 \beta_R^2} \,, \qquad
\qquad 
\langle T_R(z)\rangle =  \langle T_L(z)\rangle \mathcal{T}_L +\langle \bar{T}_R(\bar{z})\rangle \mathcal{R}_R \,.
\end{align}
Using \eqref{mainresult}, we conclude that the energy flux through the wire is given by
\beq
\langle T^{tx}\rangle=-\frac{1}{2\pi}\left(\langle T_L(z)\rangle- \langle\bar{T}_L(\bar{z})\rangle \right)= 
-\frac{1}{2\pi}\left(\langle T_R(z)\rangle- \langle\bar{T}_R(\bar{z}) \rangle\right) = 
\frac{\pi c_{LR}}{12  } \left( \frac{1}{  \beta_L^2} - \frac{  1}{ \beta_R^2} \right)\,.
\eeq
Remarkably, the heat current is still proportional to the difference of the squares of the temperatures of thermal reservoirs as in the case \eqref{currentnodefect} without interface. This fact has been noticed before in examples \cite{Bernard:2014qia}. Here, we see that it is true in general  as long as the stress tensor is the unique spin $2$ and dimension $2$ quasi-primary operator. It would be interesting to study the case where extended symmetries are present, but we leave this for future work.

\section{Conclusion}
\label{sec:summary}

In this work, we studied the scattering of conformal matter against an impurity, in 1+1 dimensions. We defined coefficients which measure the fraction of energy reflected and transmitted across the conformal interface. 
 For two CFTs whose maximal symmetry algebra is Virasoro, the reflection and transmission coefficients depend on the central charges and on the single additional coefficient $c_{LR}$, which is seen to provide a universal measure of the transparency of a conformal interface. There is no dependence on the initial momentum of the perturbation, and no dependence on the  local operator used to perturb the vacuum.  The coefficient $c_{LR}$ can be extracted from the two-point function of the stress tensor, together of course with the central charges -- see eq. \eqref{TT}.
 
The construction easily generalizes to other conserved quantities, and as an example we described the case of a global symmetry. 
One may also study the reflection and transmission coefficients of the conserved charges associated to dilations or special conformal transformations.
These are naturally defined using flux operators as in \eqref{energyOperators} with a factor of $z$ or $z^2$ inserted inside the integrals for $\Eop$ and  $\bar{z}$ or $\bar{z}^2$ for $\bEop$. Remarkably, the reflection and transmission coefficients for these charges are exactly the same as the ones for energy.
This follows from integrating    eq. \eqref{OOTxizero} times $z$ or $z^2$ over   $\int_{-\infty}^\infty dz$. 

Positivity of the reflected and transmitted energy bounds the coefficient of the two-point function of the displacement operator, \eqref{CDbound}, and with it the dependence of the free energy of the defect CFT on a small deformation of the interface. The same bound can be reinterpreted as a bound on the coefficient $\quellaR$, which was dubbed reflection coefficient in \cite{Quella:2006de}, and is indeed an average of the reflection coefficients for matter incoming from the left and from the right. 

The reflection and transmission coefficients capture a richer physics when one of the CFTs possesses an extended symmetry algebra. Carriers with different charges scatter differently against the impurity. Correspondingly, the pattern of energy reflection is captured by a larger set of CFT data, which give a more detailed description of the transparency of the interface than the sole coefficient $c_{LR}$. An example of this general situation is described in subsection \ref{subsec:cosets}.

So far we have considered interfaces betweens parity invariant CFTs. However, it is not hard to generalize our results to parity violating CFTs. If $c_L\neq \bar{c}_L$ and $c_R\neq \bar{c}_R$, then also $c_{LR} \neq \bar{c}_{LR}$, and the transmission coefficient for matter incoming from the right is modified to $\ourT_R = \bar{c}_{LR}/\bar{c}_R$, while $\ourT_L = c_{LR}/c_L$ remains unchanged. Notice, however, that a conformal interface between CFT$_L$ and CFT$_R$ only exists if the anomalies match: $c_L-\bar{c}_L=c_R-\bar{c}_R$ \cite{Billo:2016cpy}. From the two-point function of the stress tensor and the gluing condition \eqref{Tgluing1}, it is  easy to see that
 $$c_{LR}-\bar{c}_{LR}=c_L-\bar{c}_L=c_R-\bar{c}_R\,.$$
In the extreme case of two chiral theories, we conclude that they can only be glued together at an interface if they have the same chirality. Moreover, their interfaces are necessarily fully transmissive or topological, as can be deduced from the vanishing of the two-point function of the displacement operator. 
In other words,  excitations of a chiral CFT are insensitive to any impurities.
If only one of the CFTs is chiral, there are two inequivalent situations. Firstly,  if the carriers of the chiral theory go towards the interface, they are fully transmitted. Excitations coming from the other side of the interface are totally reflected. Secondly,  if instead the carriers of the chiral theory move away from the interface, their reflection and transmission coefficients are not defined. The excitations coming from the other side of the interface can be partially reflected and transmitted.

Various lines of research remain open for the future. It would be interesting to fully explore reflection and transmission in theories with extended symmetries. Rational interfaces exist in these models \cite{Quella:2002ct}, where the coefficients are computable. 

The coefficients defined in this paper only contain averaged information. It is worth looking at higher moments of the energy distribution at infinity, which may provide more detailed information on the nature of the interface and of the theories on the two sides. 

Our reflection and transmission coefficients are inclusive quantities: it would be interesting to try and define observables related to exclusive processes. We made a few remarks on this kind of observables in subsection \ref{subsec:scattering}. 
In this context, it might be fruitful to study our interface setup using the description of CFT in terms of massless particles with integrable scattering \cite{Fendley:1993wq, Bazhanov:1994ft}.
A related  direction is the study of reflection and transmission of energy in theories deformed away from the fixed point. It would be interesting to clarify the interplay between the reflection and transmission amplitudes for the particles in the  (possibly integrable) massive theory and the reflection and transmission coefficients that we  studied at the UV fixed point. It would also be important to check the role of the two-point function of the stress tensor in connection to reflection and transmission, in the deformed theory \cite{Brunner:2015vva}.

It is also natural to tackle reflection and transmission processes in higher dimensional CFTs. In dimension higher than two, there is a richer variety of defects: similar techniques to the ones developed here should be useful not only in studying the transparency of an interface, but also scattering problems involving heavy classical static particles, or sources of other codimension.

Finally, the bounds on the reflection and transmission coefficients, while having a convincing physical origin, technically rely on the assumption that the averaged null energy condition (ANEC) is true in the Hilbert space built on top of the impurity. 
It was recently noticed \cite{Lemos:2017vnx,Jensen:2018rxu} that validity of the ANEC in the vacuum in the presence of a conformal defect implies positivity 
of the one-point function of the stress tensor.
 It would be important to fill this gap and prove the ANEC in a generic defect CFT.

\section*{Acknowledgments}

We would like to thank Lorenzo Bianchi, Anatoly Konechny, Andy O'Bannon, Zohar Komargodski, Leonardo Rastelli, Balt van Rees, Brandon Robinson, Slava Rychkov and Ingo Runkel for useful discussions. MM and JP are supported by the Simons Foundation grant 488649 (Simons collaboration on the Non-perturbative bootstrap) 
and  by the Swiss National Science Foundation through the project  200021-169132 and through the National Centre of Competence in Research
SwissMAP.

\appendix

\section{Conventions}
\label{app:conventions}

Our metric convention is such that
\beq
v^2<0,\quad v\ \textup{timelike.}
\eeq
Lightcone coordinates are defined as follows:
\begin{align}
&z= x-t~, &  p=p^1-p^0~, \\
&\bz= x+t~, &  \bp=p^1+p^0~,
\end{align}
so that
\beq
p \cdot x = \frac{1}{2} (p \bz+\bp z)~,
\eeq
and 
\beq
d^2 x = \frac{1}{2} dz d\bz~.
\eeq
The $\ii \epsilon$ convention is as follows: \emph{operators at larger imaginary time stand on the right.} Therefore, a correlator of operators $O_1O_2$ ordered as written depends on the differences $z_{12}+\ii \epsilon$, $\bar{z}_{12}-\ii \epsilon$, $z_1+\bz_2+\ii\ep$ and $\bz_1+z_2-\ii \ep$, the last two being relevant in the presence of an interface.

For the stress tensor, we stick to the usual $2d$ convention, such that the OPE with a Virasoro primary operator reads
\beq
T(z) O(w,\bar{w}) \sim \frac{h}{(z-w)^2} O(w,\bar{w})+\frac{1}{z-w} \partial O(w,\bar{w}) +\textup{regular}~.
\label{TOOPE}
\eeq
$T(z)$ and $\bar{T}(\bz)$ are related as follows to the response to a deformation of the metric:
\beq
T^{\mu\nu} = -\frac{2}{\sqrt{g}} \frac{\de S}{\de g_{\m\n}}~, \qquad 
T(z)=-2\pi T_{zz}(z)~,
\eeq
and similarly for $\bar{T}$.

The OPE of a current $j^a(z)$ with a primary operator under the affine symmetry is
\beq
j^a(z) O(w,\bar{w}) \sim \frac{1}{z-w} f^a(O)(w,\bar{w}) +\textup{regular}~,
\eeq
where $f^a(O)$ is the action of the algebra element $j^a$ in the appropriate representation.

Correlation functions in a translational invariant vacuum are denoted as $\braket{\ }$, while correlation functions in the presence of an interface have a subscript: $\braketi{\ }$. When we compute overlaps in canonical quantization, we still write only one subscript: $\bra{\ }\keti{\ }$. 

The interface always extends along the time direction. When Wick rotating to Euclidean signature, this naturally places the interface on the imaginary axis. This is slightly unconventional. Our setup is related to the standard one by a rotation of $\pi/2$: $z\to \ii z$. Therefore, our formulae are obtained from the analogous ones where the interface lies on the real axis, by rotating every operator: $O \to \ii^{\bh-h} O$.

The two-point function of a scaling operator is normalized as follows
\beq
\braket{O^\dagger(z,\bz)O(0,0)}=\frac{e^{\ii \pi(h-\bh)}}{z^{2h}\bz^{2\bh}}~.
\eeq
The phase is easily justified by evaluating the correlator at $(z,\bz)=\ii(\tau,-\tau)$ and requiring it to be positive -- the cut of $z^{2h}$ is as usual on the negative real axis.

\section{Kinematics}
\label{app:kinematics}

In this appendix, $O$ is always a quasi-primary, and the signature is Euclidean, with $z=x+\ii\tau$ and the interface located at $x=0$. $O$ can acquire a one-point function in the presence of a conformal interface if it is a scalar:
\beq
\braketi{O_{h,\bh}(z,\bz)} = \frac{a_O\, \delta_{h,\bh}}{(z+\bz)^{2h}}~.
\eeq

A two-point function in which at least one operator is purely (anti)holomorphic is also fixed up to a coefficient:
\beq
\braketi{O_1(z_1,\bz_1)O_2(z_2)} = 
\frac{b_{12}}{z_{12}^{h_1+h_2-\bh_1}(z_1+\bz_1)^{h_1-h_2+\bh_1}(\bz_1+z_2)^{-h_1+h_2+\bh_1}}~.
\label{OOholo}
\eeq
If $O_2$ is on the opposite side of the interface with respect to $O_1$, then the correlator vanishes unless $h_1-\bh_1-h_2=0:$
\beq
b_{12}=0~, \qquad \textup{if} \ \ \Re z_1 \Re z_2 <0\ \ \ \textup{and}
\ \ \ h_1-\bh_1-h_2\neq 0~.
\label{b12constraint}
\eeq
This is easily seen by mapping the configuration on a sphere, with the interface on the equator and the two quasi-primaries at the two poles. Rotational invariance then forces the spin of the operators to match.
The constraint also avoids the unphysical singularity at $\bz_1=-z_2$, \emph{i.e.} when the operators are in mirroring positions with respect to the interface.

As a special case of eq. \eqref{OOholo}, when both operators are (anti)holomorphic, the correlator requires the non vanishing weights to be equal. For instance,
\beq
\braketi{\bar{O}_1(\bz_1)O_2(z_2)}=\frac{b_{12}\, \delta_{\bh_1,h_2}}{(\bz_1+z_2)^{2\bh1}}~.
\label{OOantiholo}
\eeq
The constraint \eqref{b12constraint} now implies that the two-point function of operators of opposite chirality and on opposite sides of the interface vanishes.

Formulae similar to eqs. \eqref{OOholo} and \eqref{OOantiholo} are valid for different combinations of holomorphic and antiholomorphic fields. In particular, the equation equivalent to \eqref{OOantiholo} for a pair of holomorphic operators implies that their two-point function is independent of the location of the interface. Therefore, if both operators are on the same side of the interface, $b_{12}$ can be computed in the translational invariant theory, and defines the normalization of the operator.

The two-point function of generic quasi-primaries contains one cross ratio:
\begin{multline}
\braketi{O_1(z_1,\bz_1)O_2(z_2,\bz_2)} =
\frac{b_{12}(\xi)}{z_{12}^{h_2-\bh_2}\bz_{12}^{\bh_1-h_1}(z_1+\bz_1)^{h_1+\bh_1}(z_2+\bz_2)^{h_2+\bh_2}(z_1+\bz_2)^{h_1-h_2-\bh_1+\bh_2}}~, \\
 \xi=\frac{z_{12}z_{\bar{1}\bar{2}}}{(z_1+\bz_1)(z_2+\bz_2)}~.
\end{multline}

The three-point function of purely (anti)holomorphic quasi-primary operators is fixed up to a single coefficient:
\beq
\braketi{O_1(z_1)O_2(z_2)\bar{O}_3(\bz_3)}=
 \frac{c_{12\bar{3}}}{(z_1-z_2)^{h_{12,3}}(z_1+\bz_3)^{h_{13,2}}(z_2+\bz_3)^{h_{23,1}}}~,
 \label{OOObar}
\eeq
with $h_{12,3}=h_1+h_2-\bar{h}_3$ and similarly for the other exponents. There is a constraint analogous to \eqref{b12constraint}:
\beq
c_{12\bar{3}}=0~,\qquad 
\begin{array}{l}
\textup{if}\ \ \Re z_3 \Re z_1 <0\ \ \textup{and}\ \ \Re z_1 \Re z_2 >0~, \\
\textup{if}\ \ \Re z_3 \Re z_1 <0\ \ \textup{and}\ \ \Re z_3 \Re z_2 >0~,\ \ \textup{and}\ \ h_{13,2} \neq 0~. \\
\end{array}
\label{c123constraint}
\eeq
This is easily proven by fusing the operators on the same side of the interface, and reducing to the constraint \eqref{b12constraint}, or the analogous one with opposite chiralities. Some constraints for higher-point functions can be obtained in the same way, but we will not need them here.

Of course, a formula similar to \eqref{OOObar} holds if the number of antiholomorphic operators is zero, two, or three. In particular, when all three operators are (anti)holomorphic, the correlator becomes independent of the distance from the interface. Therefore, if the operators are all on the same side, the coefficient $c_{123}$ equals the OPE coefficient as in a translational invariant theory.

When some of the operators have both holomorphic and anti-holomorphic dimensions, the three-point function depends on cross ratios. In particular, we are interested in the coupling of two generic quasi-primaries with a conserved (higher spin) current, in the presence of the interface. In this case, there are two independent cross ratios. The following representation is useful in subsection \ref{subsec:nonholo}:
\begin{multline}
\braketi{O_1(z_1,\bz_1)O_2(z_2,\bz_2)O_3(z_3)}= \\
\left(\frac{(z_2+\bz_2)(z_1-z_3)}{(z_1+\bz_1)(z_2-z_3)}\right)^{\bh_1-\bh_2}\!\!\!\!\!\!
 \frac{f_{123}(\xi,u)}{(z_1-z_3)^{h_1-h_2+h_3}(z_2-z_3)^{h_2-h_1+h_3}(z_1-z_2)^{h_1+h_2-h_3}(\bz_1-\bz_2)^{\bh_1+\bh_2}}~,\\
 \xi=\frac{z_{12}z_{\bar{1}\bar{2}}}{(z_1+\bz_1)(z_2+\bz_2)}~,\ 
 u=\frac{z_{12}(\bz_1+z_3)}{(z_1+\bz_1)z_{32}}~,
 \label{OOOgeneral}
\end{multline}

\section{Stress tensor insertions: constraints from the gluing conditions}
\label{app:OOT}

In this appendix, the defect CFT is taken to be in Euclidean signature, the setup is the same as in subsection \ref{subsec:dcft}. Consider the set of correlation functions with one stress tensor and an arbitrary number of other insertions, all of which on the left of the interface. We use the following notation:
\beq
G_{L/R}(z)=\braketi{O_1(z_1,\bz_1)\dots O_n(z_n,\bz_n) T_{L/R}(z)}~,
\eeq
while $\bar{G}_{L/R}$ denotes the analogous correlator with the insertion of $\bT_{L/R}$.
Using Cauchy theorem, or equivalently, the fact that the stress tensor is a descendant of the identity, one easily obtains the following equations
\begin{align}
G_L(z)=-\sum_i\,\textup{Res}_{w=z_i}\!\left[\frac{G_L(w)}{w-z}\right] 
+\frac{1}{2\pi \ii} \int_{-\ii\infty}^{+\ii\infty}\!dw\,
\frac{\bar{G}_L(-w)+G_R(w)-\bar{G}_R(-w)}{w-z}~, \label{GLres}\\
\bar{G}_L(\bz)=-\sum_i\,\textup{Res}_{\bar{w}=\bar{z}_i}\!
\left[\frac{\bar{G}_L(\bar{w})}{\bar{w}-\bz}\right] 
+\frac{1}{2\pi \ii} \int_{-\ii\infty}^{+\ii\infty}\!d\bar{w}\,
\frac{G_L(-\bar{w})-G_R(-\bar{w})+\bar{G}_R(\bar{w})}{\bar{w}-\bz}~. \label{GLbarRes}
\end{align}
In the second term in each line, we used the gluing condition \eqref{gluingT}. The last integrals can be performed by closing the contour always in the physical half-plane, \emph{i.e.} such that $T_L$ and $\bT_L$ are evaluated on the left of the interface and vice versa for $T_R$ and $\bT_R$ -- notice that we treat $\bz$ as a holomorphic coordinate. Then $G_R$ does not contribute to eq. \eqref{GLres} and $\bar{G}_R$ to eq. \eqref{GLbarRes}. On the other hand, the poles of $G_L$ and $\bar{G}_L$ are fixed by the OPE. Finally:
\begin{align}
G_L(z)+\bar{G}_R(-z)=-\sum_i\,\textup{Res}_{w=z_i}\!\left[\frac{G_L(w)}{w-z}\right] -\sum_i\,\textup{Res}_{w=-\bz_i}\!\left[\frac{\bar{G}_L(-w)}{w-z}\right] ~, \label{GLGR}\\
\bar{G}_L(\bz)+G_R(-\bz)=-\sum_i\,\textup{Res}_{\bar{w}=-z_i}\!\left[\frac{G_L(-\bar{w})}{\bar{w}-\bz}\right] -\sum_i\,\textup{Res}_{\bar{w}=\bz_i}\!\left[\frac{\bar{G}_L(\bar{w})}{\bar{w}-\bz}\right] ~. \label{barGLGR}
\end{align}
Let us stress that the r.h.s. of eqs. \eqref{GLGR} and \eqref{barGLGR} is fixed by the Ward identities of the stress-tensor in terms of the $n$-point function without the stress tensor insertion. The precise form of this relation depends whether the operators are primaries or descendants. 

Finally, let us briefly consider the two limiting cases of a topological and of a factorizing interface, defined by the conditions \eqref{topological} and \eqref{factorizing} respectively. If the interface is topological, the correlation function is the same as in the absence of the interface:
\begin{subequations}
\begin{align}
G_L(z) &= G_R(z) = -\sum_i\,\textup{Res}_{w=z_i}\!\left[\frac{G_L(w)}{w-z}\right]~, \\
\bar{G}_L(\bz) &= \bar{G}_R(\bz) = -\sum_i\,\textup{Res}_{\bar{w}=\bz_i}\!\left[\frac{\bar{G}_L(\bar{w})}{\bar{w}-\bz}\right]~.
\end{align}
\label{GGGGtopo}
\end{subequations}
If the interface is factorizing, we easily find that eqs. \eqref{GLGR} and \eqref{barGLGR} are complemented by 
\beq
G_R(z) = \bar{G}_R(\bz) =0~.
\label{GGGGbound}
\eeq
This is consistent with the general fact, explained in subsection \ref{subsec:dcft}, that no correlation exists between the two theories if the stress tensors on the interface obey eq. \eqref{factorizing}.

\section{The three-point function $\braketi{O_LO_LO_R}$: the $O_L\times O_L$ OPE}
\label{app:OOTOPE}

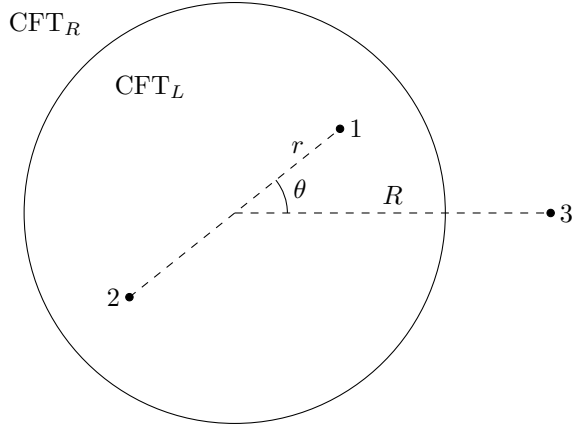
\begin{figure}
\centering
\begin{tikzpicture}[scale=1.4]
\draw [black] (0,0) circle [radius=2];
\draw [dashed] (-1,-0.8) -- node[pos=0.8,above] {$r$} (1,0.8);
\draw [dashed] (0,0) -- node[above] {$R$} (3,0);
\filldraw [black] (1,0.8) circle [radius=1pt] node[right] {$1$};
\filldraw [black] (-1,-0.8) circle [radius=1pt] node[left] {$2$};
\filldraw [black] (3,0) circle [radius=1pt] node[right] {$3$};
\draw (0.5,0) arc [start angle=0, end angle=38.66, radius=0.5] node[pos=0.7,right=1] {$\theta$};
\node at (-0.8,1.2) {CFT$_L$};
\node at (-1.8,1.8) {CFT$_R$};
\end{tikzpicture}
\caption{The configuration which defines the $\rho$-coordinates for the correlator $\braketi{O_{1,L}O_{2,L}O_{3,R}}$. The defect is a sphere with unit radius, and the points $1$ and $2$ are aligned with its center, both at distance $r<1$ from it. The right operator lies at distance $R\geq 1$ from the center of the spherical defect. The $\rho$-coordinates are the complex coordinates of the operator $O_{1,L}$, $\rho=r e^{\ii \theta}$, together with the positive real parameter $R$.}
\label{fig:rhoEuclidean}
\end{figure}

\begin{figure}
\centering
\begin{tikzpicture}[scale=1.4]
\filldraw [orange!50] (-3,-2) -- (-1,-2) -- (1,2) -- (-3,2) -- cycle;
\draw [decorate,decoration={snake,amplitude=0.8pt}] (0,0) -- (2,0);
\filldraw [black] (0,0) circle [radius=1pt] node[left] {$1$};
\draw [black] (1.9,1.8)-- (1.7,1.8) -- (1.7,2);
\node at (1.8,1.9) {$u$};
\end{tikzpicture}
\caption{The $u$ plane at fixed $\xi=4 r^2/(1-r^2)^2$, for some choice of $r$ with $|r|<1$. The region of convergence of the $O_L\times O_L$ OPE is shaded in orange. The boundary is the straight line $\Re u+\alpha\, \Im u-1=0,$ with $\al=2\, \Im r^2/(1-|r|^4)$.}
\label{fig:OPEconvergence}
\end{figure}
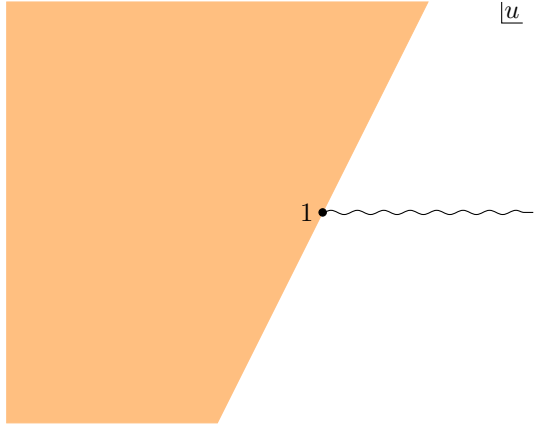

Consider the correlator of two quasi-primaries $O_{1,L},\,O_{2,L}$ of the CFT$_L$ and a quasi-primary $O_{3,R}$ of the CFT$_R$:
\beq
\braketi{O_{1,L}O_{2,L}O_{3,R}}~.
\eeq
We are interested in the region of convergence of the $O_L\times O_L$ OPE in complex cross ratio space. We follow a standard strategy, first analyzing the problem in Euclidean signature.
The correlator has three cross ratios \cite{Lauria:2017wav}, and it is not hard to see that we can use conformal transformations to place the operators and the defect in the configuration shown in fig. \ref{fig:rhoEuclidean}. We can take the cross ratios to be the complex coordinates $(\r,\rb)$ of the operator $O_{1,L}$, together with the distance $R$ of $O_{3,R}$ from the origin:
\beq
\textup{cross ratios:} \qquad \quad \r=\rb^*=r e^{\ii \theta}~, \quad R~.
\label{rhocoord}
\eeq
These cross ratios have the following crucial property: the region of convergence of the Euclidean OPE equals the radius of convergence of the expansion of the correlator in powers of $r^2=\r\rb$, \emph{i.e.} $r<1$. This kind of $\r$-coordinates were introduced for the four-point function of local operators in \cite{Pappadopulo:2012jk}, studied in \cite{Hogervorst:2013sma} and later extended to other contexts, including the domain of defect CFTs \cite{Lauria:2017wav}. The power expansion of the correlator reads
\beq
\braketi{O_{1,L}(\rho)O_{2,L}(-\rho)O_{3,R}(R)}= (2\rho)^{-h_1-h_2}(2\rb)^{-\bh_1-\bh_2} 
\sum_{h,\bh\geq0} a_{h,\bh}(R) \r^h\rb^{\bh}~, \qquad \rb=\r^*~,
\eeq
where it is important that the powers of $\r$ and $\rb$ are bounded from below. Although the coefficients $a_{h,\bh}(R)$ are not positive, the expansion converges absolutely \cite{Pappadopulo:2012jk}. It is then easy to show that the following is true for independent complex $(\r,\rb)$:\footnote{The proof goes as follows: start with $|\rb|\leq|\r|<1$. Then the expansion is bounded by the following chain of inequalities:
\beq
\left|\sum_{h,\bh\geq0} a_{h,\bh}(R) \r^h\rb^{\bh} \right| \leq
\sum_{h,\bh\geq0} |a_{h,\bh}(R)||\r^h||\rb^{\bh}| \leq
\sum_{h,\bh\geq0}| a_{h,\bh}(R)| |\r|^{h+\bh} <\infty~.
\eeq
The last inequality follows from the absolute convergence of the Euclidean OPE. Exchanging $\r$ and $\rb$ the general result is obtained.}
\beq
O_{1,L}\times O_{2,L} \ \ \textup{OPE converges if}\qquad |\r|<1 \land |\rb|<1~,\ \forall R\geq 1~.
\label{OPEdomain}
\eeq
Let us now specialize to the case where $O_{3,R}$ is holomorphic, which is relevant for the transmission coefficient, see eq. \eqref{OOTR}. The relation of the $\r$ coordinates with $\xi$ and $u$ is
\beq
\xi = \frac{4\r \rb}{(1-\r\rb)^2}=\frac{4\w \wb}{(1-\w\wb)^2}~,\qquad u=\frac{2\r (1-R\rb)}{(\r+R)(1-\r\rb)}=\frac{2\w (1-\wb)}{(1+\w)(1-\w\wb)}~.
\label{omegatoxi}
\eeq
We introduced the combinations $\w=\r/R$ and $\bar{\w}=\rb R$, consistently with the fact that only two of the three cross ratios survive. We are interested in the intersection of the domain \eqref{OPEdomain} with the complex $u$ plane at fixed $\xi$. This is obtained as follows. Fix a complex $r^2=\w\wb$, thus fixing $\xi$, and consider the image of the domain $r^2/R<|\w|<1/R$ under the map \eqref{omegatoxi}. Then vary $R\geq 1$ and take the union: this is the largest region of convergence of the OPE in the $u$ plane. Obviously, the union of the domains is the region in the $u$ plane corresponding to $0<|\w|<1$. The result is shown in fig. \ref{fig:OPEconvergence}. Notice that one of the boundaries ($|\w|=1$) passes through $u=1$ and $u=\infty$, which correspond to the Lorentzian configurations where the $O_{3,R}$ is lightlike separated from either $O_{1,L}$ or $O_{2,L}$ -- see fig. \ref{fig:spacetimecontour}. The other boundary is the point $u=-2 r^2/(1-r^2)$, which corresponds to $\w=0$: here the OPE in fact converges. Indeed, this point can be reached by choosing $R=\infty$, at fixed $|\r|,\,|\rb|<1$.
To summarize, \emph{the region of convergence of the $O_L\times O_L$ OPE in the correlator $\braketi{O_{1,L}O_{2,L}O_{3,R}}$ where $O_{3,R}$ is holomorphic is $|\w|<1\ |\w\wb|<1$.}

\section{Details on $\braketi{\si_L|\Eop_R|\si_L}$ in the Ising model}
\label{app:ising}

This appendix gathers some details on the computations performed in subsection \ref{subsec:ising}.

\begin{figure}[]
\centering
\begin{tikzpicture}[scale=1.2]
\filldraw [black] (-1,1) circle [radius=1pt] node[above] {$z_1$};
\filldraw [black] (-1.5,-1) circle [radius=1pt] node[above] {$z_2$};
\node at (0.15,1.8) {$w$};
\draw [dashed,postaction = {decoration={markings, mark=between positions 0.2 and 0.9 step 1cm  with {\arrow{>};}},decorate}] (0,-2) -- (0,-0.3);
\draw [dashed,postaction = {decoration={markings, mark=between positions 0.2 and 0.9 step 1cm  with {\arrow{>};}},decorate}] (0,0.3) -- (0,2);
\draw [dashed,postaction = {decoration={markings, mark=between positions 0.2 and 0.9 step 1cm  with {\arrow{>};}},decorate}] (0,-0.3) -- (-2,-0.3);
\draw [dashed,postaction = {decoration={markings, mark=between positions 0.2 and 0.9 step 1cm  with {\arrow{>};}},decorate}] (-2,0.3) -- (0,0.3);
\draw [postaction = {decoration={markings, mark=between positions 0.2 and 0.9 step 1cm  with {\arrow{>};}},decorate}] (-2,0) -- (2,0);
\draw [black] (3,1.8)-- (2.8,1.8) -- (2.8,2);
\node at (2.92,1.91) {$z$};
\end{tikzpicture}
\caption{Contours which define the expectation value $\braketi{\si_L|\Eop_R|\si_L}$ at order $g$. The solid horizontal line is the contour of integration in $z$, while the dashed line is the contour of integration in $w$. The integrand, eq. \eqref{IsinggIntegrand} has poles in $z$ at $z=z_1,\,z_2,\,w$. The position of $z_1$ and $z_2$ is compatible with the kinematics described in subsection \ref{subsec:nonholo}.}
\label{fig:isinggz}
\end{figure}
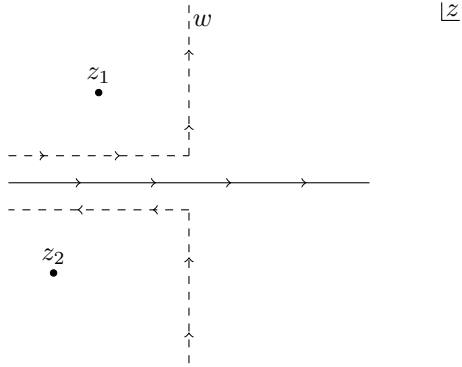

We start by proving that the transmission coefficient, as computed from $\braketi{\si_L|\Eop_R|\si_L}$, vanishes at order $g$. We need to compute -- see eq. \eqref{energyOperators}:
\beq
-\frac{1}{2\pi}\int_{-\infty}^{+\infty}\!\! dz\, G_1(z_1,z_2,\bz_1,\bz_2,z)~.
\eeq
$G_1$ was defined in subsection \ref{subsec:ising}.
The kinematics in this case is the one explained in subsection \ref{subsec:nonholo}: $z_1+\bz_1$ and $z_2+\bz_2$ are real, and $\Im z_1>0>\Im z_2$. It is convenient to perform the $z$ integration first. We need to integrate the stress tensor on the real axis, but the contour of integration in $w$ must be kept on the left of $z$. Hence, we first deform the $w$ contour as in fig. \ref{fig:isinggz}.
By first closing the $z$ contour, say, above, we pick up a contribution from the pole at $z=z_1$, and a contribution from the pole at $z=w$ in eq. \eqref{IsinggIntegrand}. The first contribution is given by the following integral:
\beq
-\frac{1}{2\pi}\int_{-\infty}^{+\infty}\!\! dz\, G_1(z_1,z_2,\bz_1,\bz_2,z) \to \frac{\pa}{\pa z_1}\int_{-\ii\infty}^{+\ii\infty}dw \braket{\si_1\si_2\ep}~.
\eeq
$\braket{\si_1\si_2\ep}$ is defined in eq. \eqref{threepointising}. Let us analyze its behavior in the $D\to \infty$ limit, \emph{i.e.} when $\bz_1\,\ \bz_2\to-\infty$ at fixed $\bz_1-\bz_2$. The contour of integration in $w$ can be freely shifted towards the right -- see fig. \eqref{fig:isingg}, so that $w$ lies at a distance at least of order $D$ from both $\bz_1,\, \bz_2$ and $z_1,\,z_2$. It is then immediate to show, looking at eq. \eqref{threepointising}, that the integral falls off as $D^{-1}$. The contribution from the pole at $z=w$ only involves an integral over the upper part of the dashed contour in fig. \ref{fig:isinggz}. We can then deform the $w$ contour on the $z_1$ branch cut of fig. \ref{fig:isingg}, and obtain the following expression
\begin{multline}
-\frac{1}{2\pi}\int_{-\infty}^{+\infty}\!\! dz\, G_1(z_1,z_2,\bz_1,\bz_2,z) \\
\to -2\ii
\left(\frac{\pa}{\pa z_1}+\frac{\pa}{\pa z_2}\right)\left\{  \frac{(z_{12}\bz_{12})^{3/8}}{|z_1+\bz_1|}
 \int_{0}^{+\infty}\!\!  \frac{dx}{\sqrt{x (x+1)}} \left(-x+\frac{z_1-z_2}{|z_1+\bz_1|}\right)^{-1/2}  
 \left(x+1+\frac{\bz_1-\bz_2}{|z_1+\bz_1|}\right)^{-1/2}\right\}~,
\end{multline}
where all the square root cuts run on the negative real axis. It is easy to see that the integral has a logarithmic behavior as $|z_1+\bz_1|\to\infty$, so that also this contribution vanishes, and we confirm that the transmission coefficient only deviates from 1 at order $g^2$.

We now turn to the subleading order. First of all, let us address a regularization issue. The double integral in eq. \eqref{IsingG2} is UV divergent, since the integrand goes like $(w_1-w_2)^{-2}$ as $w_1 \to w_2$. We regulate with a hard cutoff, and ignore power-law divergences, which is equivalent to the following prescription:
\beq
\int_{-\ii \infty}^{\ii\infty}\!\! dw_1 dw_2 \to 
\int_{-\ii \infty}^{\ii\infty}\!\! dw_1 \left(\int_{-\ii \infty+\eta}^{\ii\infty+\eta} dw_2-\frac{1}{2} \oint_{w_1} dw_2\right)~,
\label{Isingreg}
\eeq
where $\eta$ is a finite positive number, and the second integral in parenthesis would pick up a simple pole at $w_2=w_1$. However, a simple pole is absent from the $\ep \times \ep$ OPE, hence we can drop the second term in parenthesis in eq. \eqref{Isingreg}. In practice, we can freely shift the two contours independently.

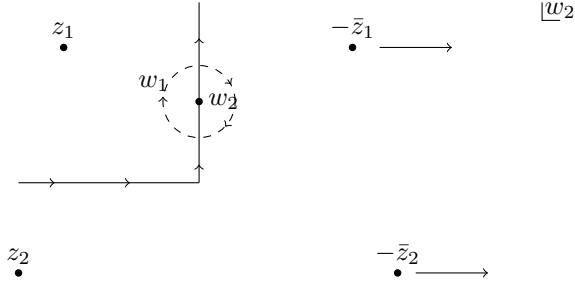
\begin{figure}
\centering
\begin{tikzpicture}[scale=1.2]
\filldraw [black] (-1.5,1.5) circle [radius=1pt] node[above] {$z_1$};
\filldraw [black] (-2,-1) circle [radius=1pt] node[above] {$z_2$};
\filldraw [black] (1.7,1.5) circle [radius=1pt] node[above] {$-\bz_1$};
\filldraw [black] (2.2,-1) circle [radius=1pt] node[above] {$-\bz_2$};
\filldraw [black] (0,0.9) circle [radius=1pt] node[right] {$w_2$};
\node at (-0.5,1.1) {$w_1$};
\draw [postaction = {decoration={markings, mark=between positions 0.1 and 1 step 0.7  with {\arrow{>};}},decorate}] (0,0) -- (0,2);
\draw [postaction = {decoration={markings, mark=between positions 0.2 and 0.9 step 1cm  with {\arrow{>};}},decorate}] (-2,0) -- (0,0);
\draw [dashed,postaction = {decoration={markings, mark=between positions 0.1 and 1 step 0.4  with {\arrow{<};}},decorate}] (0,0.9) circle [radius=0.4];
\draw[->] (2,1.5) -- (2.8,1.5);
\draw[->] (2.4,-1) -- (3.2,-1);
\draw [black] (4,1.8)-- (3.8,1.8) -- (3.8,2);
\node at (4,1.91) {$w_2$};
\end{tikzpicture}
\caption{Contours of integration responsible for the value of $\ourT_L$ at order $g^2$. The integrand is proportional to $\textup{Res}_{z=w_2} \braket{\si(z_1,\bz_1)T(z)\si(z_2,\bz_2)\ep(w_1,-w_1)\ep(w_2,-w_2)}$. The dashed circle is the contour of integration in $w_1$, which picks up the pole at $w_1=w_2$. Finally, the solid line is the contour of integration in $w_2$. The singularities at $w_2= z_1,\,z_2,\,-\bz_1,\,-\bz_2$ are poles for the final integrand, even before taking the $D\to\infty$ limit, whose effect on the position of $\bz_1$ and $\bz_2$ is illustrated by the arrows.}
\label{fig:isingt1t2}
\end{figure}

We showed in the main text that the correlator has the limiting behavior prescribed in eq. \eqref{OOTxizero}.
Here we also verify that the integration over $z$ commutes with the $D \to \infty$ limit. In other words, we would like to compute
\beq
-\frac{1}{2\pi}\lim_{D\to\infty} \int_{-\infty}^{+\infty}\!\!dz\, G_2(z_1,z_2,\bz_1-D,\bz_2-D,z)~,
\eeq
where $G_2$ was defined in subsection \ref{subsec:ising}. As before, we first perform the $z$ integral. The deformation of the $w_1$ and $w_2$ contours in the first line of eq. \eqref{IsingG2} proceeds like in fig. \ref{fig:isinggz}. Then we close the $z$ contour, say, up. It is not hard to see that the only contribution which does not vanish in the $D\to\infty$ limit, or does not cancel against the second line in eq. \eqref{IsingG2}, comes from picking the pole at $z=w_2$, where the $w_2$ contour is the right-most one -- see eq. \eqref{Isingreg}. Then we shift the $w_1$ contour to the right, and in doing so we pick up the pole at $w_1=w_2$. The final integration contours in $w_1$ and $w_2$ is illustrated in fig. \ref{fig:isingt1t2}. 
At this point, the $D\to\infty$ limit can be safely taken inside the integral, and the result is
\begin{multline}
-\frac{1}{2\pi}\left(\frac{g}{2\pi}\right)^2\lim_{D\to\infty} \int_{-\infty}^{+\infty}\!\!dz\, G_2(z_1,z_2,\bz_1-D,\bz_2-D,z) 
= \ii g^2 \textup{Res}_{w_2=z_1} \frac{1/16}{\bz_{12}^{1/8}z_{12}^{-15/8}(z_1-w_2)^2(z_2-w_2)^2} \\
= \frac{g^2}{2\pi} \int_{-\infty}^{\infty}\!\! dz\, \braket{\si(z_1,\bz_1)T(z)\si(z_2,\bz_2)}~.
\end{multline}
This is compatible with the integral of both sides of eq. \eqref{isingDinfty}, and thus confirms that the transmission coefficient can be computed by first taking the $D\to\infty$ limit and then integrating the stress tensor.

\bibliography{./auxi/bibliography}
\bibliographystyle{./auxi/JHEP}

\end{document}